\documentclass[twoside]{article}

\usepackage{aas2pp4,epsfig}

\def\eg{{e.g.,}\ }
\def\ie{{i.e.,}\ }
\def\kms{{\rm\,km\,s^{-1}}}

\begin{document}

\lefthead{KOEKEMOER \& BICKNELL}
\righthead{RADIO GALAXY EMISSION-LINE REGIONS: PKS 2356$-$61}

\title{Dynamics and Excitation of Radio Galaxy Emission-Line Regions
	--- I. PKS 2356$-$61}

\author{Anton M. Koekemoer}
\affil{Space Telescope Science Institute, 3700 San Martin Drive, Baltimore,
	MD 21210, USA; koekemoe@stsci.edu}

\author{Geoffrey V. Bicknell}
\affil{ANU Astrophysical Theory Centre, Australian National University,
	Canberra, ACT 0200, Australia
	(The ANUATC is operated jointly by the Mount Stromlo and Siding Spring
	Observatories and the School of Mathematical Sciences of the Australian
	National University); gvb@maths.anu.edu.au}

\begin{abstract}
Results are presented from a programme of detailed longslit spectroscopic
observations of the extended emission-line region (EELR) associated with the
powerful radio galaxy PKS~2356$-$61. The observations have been used to
construct spectroscopic datacubes, which yield detailed information on the
spatial variations of emission-line ratios across the EELR, together with its
kinematic structure. We present an extensive comparison between the data and
results obtained from the MAPPINGS~II shock ionization code, and show that
the physical properties of the line-emitting gas, including its ionization,
excitation, dynamics and overall energy budget, are entirely consistent with a
scenario involving auto-ionizing shocks as the dominant ionization mechanism.
This has the advantage of accounting for the observed EELR properties by means
of a single physical process, thereby requiring less free parameters than the
alternative scheme involving photoionization by radiation from the active
nucleus. Finally, possible mechanisms of shock formation are considered in the
context of the dynamics and origin of the gas, specifically scenarios involving
infall or accretion of gas during an interaction between the host radio galaxy
and a companion galaxy.
\end{abstract}


\vspace{0.5in}
\received{1997 June 26}
\accepted{1997 November 7}
\sluginfo

{\centering To appear in {\it The Astrophysical Journal}\par}

\section{Introduction}

Powerful radio galaxies are often observed to contain systems of ionized gas,
extended up to scales of several tens of kpc from the nucleus. Detailed studies
of these extended emission-line regions (EELRs) have revealed a wide range of
morphological and kinematic properties
(\eg Baum \& Heckman 1989a, 1989b; Tadhunter, Fosbury \& Quinn 1989;
McCarthy, Spinrad, \& van Breugel 1995):
\nocite{Baum.1989.ApJ.336.681,
	Baum.1989.ApJ.336.702,
	Tadhunter.1989.MNRAS.240.225,
	McCarthy.1995.ApJS.99.27}
%
%
in some objects the EELR is coincident with the radio lobes, while in the
majority of cases the line-emitting gas is extended at modest angles away from
the radio axis and does not display a direct spatial association with the radio
emission. The EELR kinematics
(Tadhunter et~al. 1989; Heckman et~al. 1989; Baum et~al. 1992)
	\nocite{Tadhunter.1989.MNRAS.240.225,
	Heckman.1989.ApJ.338.48,
	Baum.1992.ApJ.389.208}
	%
	%
can be generally classified into ``rotators'' and ``non-rotators'', the latter
being further subdivided into ``calm'' and ``violent''. The rotator EELRs
typically exhibit circular velocity fields approximately consistent with the
massive halos inferred for radio ellipticals, while the non-rotators display
less well-ordered kinematics and little or no evidence of settled dynamics.

Explanations for the origin of the gas include mergers or interactions
involving gas-rich galaxies; large differences between the gaseous and stellar
kinematics
(Phillips et~al. 1986; Tadhunter et~al. 1989; Bettoni et~al. 1990)
\nocite{Phillips.1986.AJ.91.1062,
	Tadhunter.1989.MNRAS.240.225,
	Bettoni.1990.AJ.99.1789}
%
%
suggest external infall of gaseous material on recent timescales compared to
the age of the elliptical
(Habe \& Ikeuchi 1985, 1988; Varnas et~al. 1987).
\nocite{Habe.1985.ApJ.289.540,
	Habe.1988.ApJ.326.84,
	Varnas.1987.ApJ.313.69}
%
%
Detailed kinematic studies of EELRs often reveal a correspondence between the
radio axis and the gas ``rotation'' axes
(\eg Simkin 1977; Heckman et~al. 1985; Tadhunter et~al. 1989),
\nocite{Simkin.1977.ApJ.217.45,
	Heckman.1985.ApJ.299.41,
	Tadhunter.1989.MNRAS.240.225}
%
%
%
suggesting a direct connection between gas infall and nuclear activity. It has
also been suggested that the gas has condensed out of a hot intergalactic
medium
(Fabian et~al. 1981; Canizares et~al. 1987);
\nocite{Fabian.1981.MNRAS.196.P35,
	Canizares.1987.ApJ.312.503}
%
%
however, the angular momentum of gas in radio galaxies tends to be higher than
that of typical cooling flow EELRs. Furthermore, the importance of mergers in
fuelling galaxy activity is suggested by the typical morphological disruption
of host galaxies
(Hutchings 1987; Smith \& Heckman 1989)
\nocite{Hutchings.1987.ApJ.320.122,
	Smith.1989.ApJ.341.658}
%
%
as well as apparent environmental differences between radio-loud and normal
ellipticals
(Lilly \& Longair 1984; Spinrad \& Djorgovski 1984).
\nocite{Lilly.1984.MNRAS.211.833,
	Spinrad.1984.ApJ.280.L9}
%

Thus, a principal question in the study of the formation and evolution of radio
galaxies concerns the detailed physical properties of the EELRs. In particular,
models of the EELR ionization and dynamics are of importance in understanding
their origin, as well as their r{\^o}le as a source of fuel for the active
nucleus, related to the more general question of the triggering mechanism for
active galaxies.

Excitation of radio galaxy EELRs has generally been considered in the context
of ``unified schemes'' of radio galaxies and quasars
(\eg Barthel 1989),
\nocite{Barthel.1989.ApJ.336.606}
%
%
involving photoionization of the gas by radiation from the active nucleus. This
is analogous to the scenario inferred on kpc scales for EELRs in Seyfert
galaxies, where nuclear photoionization is supported by ``ionization cones'' in
line-ratio maps in a number of objects, and by the presence of polarized broad
emission lines
(\eg Antonucci \& Miller 1985; Mulchaey, Wilson, \& Tsvetanov 1996).
\nocite{Antonucci.1985.ApJ.297.621,Mulchaey.1996.ApJS.102.309}
%
%
While evidence for orientation-based unification of radio galaxies and quasars
has been inferred from a range of observational studies, including radio galaxy
morphology and beaming studies, X-ray properties, and host galaxy populations
(\eg Antonucci 1993 and references therein;
see also Gopal-Krishna, Kulkarni, \& Wiita 1996),
\nocite{Antonucci.1993.ARAA.31.473,
	Gopal-Krishna.1996.ApJ.463.L1}
%
%
it is less clear that all the detailed properties of the line-emitting gas in
all radio galaxies can be explained in terms of nuclear photoionization.
Firstly, radio galaxy EELRs are extended on much larger scales than those in
Seyfert galaxies, typically by at least an order of magnitude, and do not
display clear ionization cones. Furthermore, relationships between EELR
luminosity and core power or total radio power typically display a scatter of
$1 - 2$ orders of magnitude
(\eg Baum \& Heckman 1989b),
\nocite{Baum.1989.ApJ.336.702}
%
%
which may be due in part to different ionization mechanisms dominating in
different objects that have otherwise similar core properties. Another
interesting effect is that the global kinematic properties of EELRs have been
shown to correlate with their overall excitation state; this is not a directly
expected consequence of unified schemes, and suggests the importance of local
ionization in some classes of objects
(Baum et~al. 1990, 1992).
\nocite{Baum.1990.ApJS.74.389,
	Baum.1992.ApJ.389.208}
%

Moreover, detailed studies of individual radio galaxies have shown that the
EELR properties cannot always be readily accounted for in terms of nuclear
photoionization. Specifically, the extended material in a number of objects
display emission-line fluxes and excitation states that are inconsistent with
the nuclear ionization properties required to photoionize the material closer
to the AGN
(\eg Danziger et~al. 1984; van Breugel et~al. 1986; Tadhunter et~al. 1988;
Meisenheimer \& Hippelein 1992).
\nocite{Danziger.1984.MNRAS.208.589,
	vanBreugel.1986.ApJ.311.58,
	Tadhunter.1988.MNRAS.235.403,
	Meisenheimer.1992.AA.264.455}
%
%
In such objects, the importance of local ionizing mechanisms, such as shocks,
has been postulated in order to explain the large-scale properties of
the extended gas.
Thus, while orientation effects are
clearly likely to affect some observational properties of radio galaxies and
quasars, we point out that the ionization and excitation properties of the
EELRs may be dominated to varying degrees in individual objects by local
ionization mechanisms, related for example to shocks.

In this paper we address in detail the feasibility of ``auto-ionizing shocks''
as a viable ionization mechanism for the extended gas. One possible origin for
shocks is collisions between streams of gas during a galaxy merger. This
scenario is motivated by the ``tidal'' EELR morphology often observed in radio
sources, where the gas is neither conical nor directly associated with the
radio plasma. In objects where the radio plasma coincides with the EELR
(\eg Prieto et~al. 1993; Tadhunter et~al. 1994)
\nocite{Prieto.1993.MNRAS.263.10,
	Tadhunter.1994.AA.288.L21}
%
%
it is possible that associated bowshocks may play a direct role in ionizing the
gas. For example, the auto-ionizing shock model has been applied successfully
to observations of emission-line gas in Centaurus~A
(Sutherland, Bicknell \& Dopita 1993).
\nocite{Sutherland.1993.ApJ.414.510}
%
%
The physical appeal of auto-ionizing shocks is that the gas energetics,
excitation {\it and} kinematics are accounted for in a single, physically
self-consistent model, requiring less free parameters than photoionization by a
hidden AGN.

\section{Observational Programme}

\subsection{Target Object}

In order to obtain as complete a description as possible of the physical
properties of the gas, a combination of kinematic and excitation information is
required across its {\it entire} extent. We describe the results from a program
of low- and high-dispersion longslit spectroscopy of the radio galaxy
PKS~2356$-$61, sampling the EELR completely through the use of multiple slit
positions.

The radio galaxy PKS~2356$-$61 is one of the strongest southern FR~II sources,
with a total radio power $P_{\rm 1.4\,GHz} \sim 10^{25.8}\,$W$\,$Hz$^{-1}$, and
was discovered during the ``Mills Cross'' radio survey of the southern sky by
Mills et~al. (1961).
\nocite{Mills.1961.AJP.14.497}
%
%
Its double-lobed morphology has been apparent since the first radio synthesis
observations were obtained
(Ekers 1969).
\nocite{Ekers.1969.AJPAS.6.1}
%
%
The optical counterpart was identified by
Westerlund \& Smith (1966)
\nocite{Westerlund.1966.AJP.19.181}
%
%
to be a type E3 or D elliptical of $V$ magnitude~16, in a sparsely populated
field (containing 10 other galaxies within a region $\sim 21\arcmin$ in size,
\ie $\sim 2.3\,$Mpc for ($H_0 = 75\kms\,$Mpc$^{-1}$, $q_0 = 0$, which we use
throughout this paper). The first redshift determination was by
Whiteoak (1972),
\nocite{Whiteoak.1972.AJP.25.233}
%
%
who measured a value of $z = 0.0959$ based on optical emission-lines. A nuclear
spectrum obtained by
Danziger \& Goss (1983)
\nocite{Danziger.1983.MNRAS.202.703}
%
%
shows strong, high-excitation narrow-line emission, with a high
[OIII]$\lambda$5007/H$\beta$ ratio ($\gtrsim 10$) and the presence of
significant [NeV]$\lambda$3426. This is also evident in the fluxes published by
Robinson et~al. (1987)
\nocite{Robinson.1987.MNRAS.227.97}
%
%
for the nuclear region and two off-nuclear regions (7\arcsec~E and 4\arcsec~W).
These and other optical properties are summarized in Table~\ref{tab_gen_props}.

\begin{deluxetable}{lrl}
\tablewidth{0pt}
\tablecaption{\label{tab_gen_props}
	General properties of PKS~2356$-$61.}
\tablehead{	&			&	Notes	}
\startdata
R.A. (B1950)	& $23^{\rm h}\;56^{\rm m}\;29\fs37\;\pm\,0\fs02$	& 1,2 \nl
Dec. (B1950)	& $-61\arcdeg\;11\arcmin\;40\farcs6\;\pm\,0\farcs5\!\phn$	& 1,2		\nl
$z_{\rm abs}$	& 0.0963~\phm{$\pm\,0.0001$}				& 3 \nl
$z_{\rm em}$	& 0.0959~$\pm\,0.0001$					& 3,4 \nl
$z_{\rm abs + em}$	& 0.0958~$\pm\,0.0003$				& 5 \nl
Distance $D_L$ $z_{\rm abs + em}$ (Mpc)$^\dag$	& 383	\hspace{3em}\mbox{}	& \nl
Scale (kpc/arcsec)$^\dag$			& 1.86	\hspace{3em}\mbox{}	& \nl
Galaxy type	& D/E3				\hspace{2.5em}\mbox{}	& 1,2,6 \nl
$V$ mag.	& 16				\hspace{3em}\mbox{}	& 1 \nl
$B$ mag.	& 17				\hspace{3em}\mbox{}	& 6,7 \nl
Continuum extent	& $6\arcsec \times 4\arcsec$	\hspace{2em}\mbox{}	& 1 \nl
Continuum P.A.		& $6\arcdeg$		\hspace{3em}\mbox{}	& 1 \nl
EELR extent		& $14\arcsec \times 8\arcsec$	\hspace{2em}\mbox{}	& 8 \nl
$F_{\rm [OIII]\lambda5007}/F_{\rm H\beta}$	& 13.2	\hspace{3em}\mbox{}	& 9 \nl
$F_{\rm H\beta}$(erg/s/cm$^2$)	& $3.94 \times 10^{-15}$ \hspace{1em}\mbox{}	& 9 \nl
$\Delta V_{\rm tot}$ (km\,s$^{-1}$)	& 180	\hspace{3em}\mbox{}	& 8 \nl
$\Delta V_{\rm max}$ (km\,s$^{-1}$)	&  90	\hspace{3em}\mbox{}	& 8 \nl
$P_{\rm 1.4\,GHz}$ (W$\,$Hz$^{-1}$)	& $6.3 \times 10^{25}$ \hspace{1em}\mbox{}	& 10 \nl
\enddata
\tablecomments{
1:~Westerlund \& Smith 1966;	\nocite{Westerlund.1966.AJP.19.181}
2:~Ekers 1970;			\nocite{Ekers.1970.AJP.23.217}
3:~Tritton 1972;		\nocite{Tritton.1972.MNRAS.158.77}
4:~Whiteoak 1972;		\nocite{Whiteoak.1972.AJP.25.233}
5:~Danziger \& Goss 1983;	\nocite{Danziger.1983.MNRAS.202.703}
6:~Sutton 1968;			\nocite{Sutton.1968.AJP.21.221}
7:~Tadhunter et~al. 1989;	\nocite{Tadhunter.1989.MNRAS.240.225}
8:~Danziger \& Focardi 1988;	\nocite{Danziger.1988.CFCG.133}
9:~Robinson et~al. 1987;	\nocite{Robinson.1987.MNRAS.227.97}
10:~Wright \& Otrupcek 1990.\\	\nocite{Wright.1990.PKSCAT90}
%
%
$^\dag$ We use $H_0 = 75\kms\,$Mpc$^{-1}$, $q_0 = 0$.
}
\end{deluxetable}

The only previously published emission-line image is in H$\alpha$
(Danziger \& Focardi 1988),
\nocite{Danziger.1988.CFCG.133}
%
%
showing a bright, elongated emission-line region centred on the galaxy nucleus,
with an arc or shell-like feature on its eastern side. A very faint tail
extends from the east to the south; a high-dispersion longslit spectrum of one
of the blobs in this tail
(Tadhunter et~al. 1989)
\nocite{Tadhunter.1989.MNRAS.240.225}
%
%
shows a relatively small velocity gradient ($\lesssim 200\kms$).

\subsection{Observations and Data Reduction}

The observations were obtained during dark time on the nights of 1 and 2
August, 1992, using the RGO Spectrograph mounted at the f/8 Cassegrain focus of
the Anglo-Australian Telescope (AAT). The detector used was the
Tektronix~\#2~CCD, which consists of $1024 \times 1024$ 24\micron\ pixels, each
corresponding to 0\farcs77 on the sky. Details of the observations are
presented in Table~\ref{tab_obsns}. Conditions were sufficiently photometric
that reliable absolute photometry was obtained, and the seeing ranged between
1\farcs2 and 1\farcs5. A detailed account of the observations and reduction
procedures is presented in
Koekemoer (1996).
\nocite{Koekemoer.1996.PhD}
%

\begin{deluxetable}{lcc}
\tablewidth{0pt}
\tablecaption{\label{tab_obsns}
	Spectrophotometric Observations of PKS~2356$-$61.}
\tablehead{		& 1 August 1992		& 2 August 1992	}
\startdata
Grating			& 270R			& 1200V			\\
Wavelength/pixel	& 3.39\,\AA		& 0.58\,\AA		\\
Spectral range		& 3996$\,\rightarrow\,$7465\,\AA	& 5305$\,\rightarrow\,$5895\,\AA	\\
Emission lines		& [OII]$\lambda$3727$\rightarrow$[SII]$\lambda$6731	& H$\beta$, [OIII]$\lambda$5007			\\
Calibration arcs	& He/Ne/Ar		& Cu/Ar			\\
\enddata
\tablecomments{
The data obtained each night consisted of exposures in 14 parallel, adjacent
slit positions at a position angle of 0\arcdeg. The slit width was 1\arcsec\ and
the slit positions were offset in R.A. in increments of 1\arcsec; the spatial
coverage extended from $5\arcsec\,$W$\,\rightarrow\,8\arcsec\,$E of the nucleus.
The exposure time at each slit position totaled 1200~sec.
}
\end{deluxetable}

Longslit exposures were obtained at parallel positions across the entire extent
of the EELR, thereby producing two spectral datacubes (at high- and
low-dispersion) in Right Ascension, Declination and wavelength.  New slit
positions were acquired by offsetting the telescope in increments of 1\arcsec.
This is most reliably achieved through the use of ``Star-to-Probe'' guiding:
the guiding assembly is moved by a specified amount, and the telescope moves
automatically in order to re-position the guide star on a standard location. By
re-acquisition of the galaxy nucleus and nearby stars throughout the night, it
was verified that positioning errors were $\lesssim 0\farcs15$.

Longslit observations are susceptible to problems due to atmospheric effects.
Seeing fluctuations can cause changes in the contribution from regions
immediately outside the slit, while low-dispersion data can suffer from
differential refraction, resulting in varying contributions from regions
outside the slit as a function of wavelength. However, all the data were
obtained at zenith distances $\lesssim 30 - 40\arcdeg$, thus differential
refraction effects are $\lesssim 0\farcs2$ along the wavelength regime covered.
The impact of these effects was further minimized by employing a slit width of
1\arcsec, comparable to the seeing disk size, and by spatially smoothing the
final dataset using 3$\times$3-pixel averaging.

Smooth-spectrum stars and photometric standards
(Baldwin \& Stone 1984)
\nocite{Baldwin.1984.MNRAS.206.241}
%
%
were observed to correct for atmospheric absorption lines and the
detector/gra\-ting spectral response. Routines in the IRAF ``onedspec'' package
were used to apply these corrections to the data. Frames were de-biased using
the corresponding overscan regions, and no two-dimensional bias structure was
present on the detector. Pixel-to-pixel flat-fields were obtained from
exposures of the internal tungsten lamp. The frames were corrected for optical
vignetting along the slit using twilight exposures obtained with the same slit
width. The spectra were corrected for atmospheric extinction using standard
extinction curves for Siding Spring Observatory (available within IRAF).

Internal arc exposures were obtained at regular intervals throughout each
night, in order to correct for instrumental shifts. A two-dimensional
wavelength transformation for each arc frame, subsequently applied to the data,
was obtained by fitting arc lines at multiple positions along the slit, using
routines in the IRAF ``twodspec'' package. The resulting residuals in
wavelength were less than 0.1 pixels across the regions of interest. After
wavelength calibration, the night-sky spectrum for each frame, obtained from
empty regions along the slit, was subtracted from the data.

\section{Results}

\subsection{Morphology}

\begin{figure*}
\begin{minipage}{\linewidth}
\centering\epsfig{file=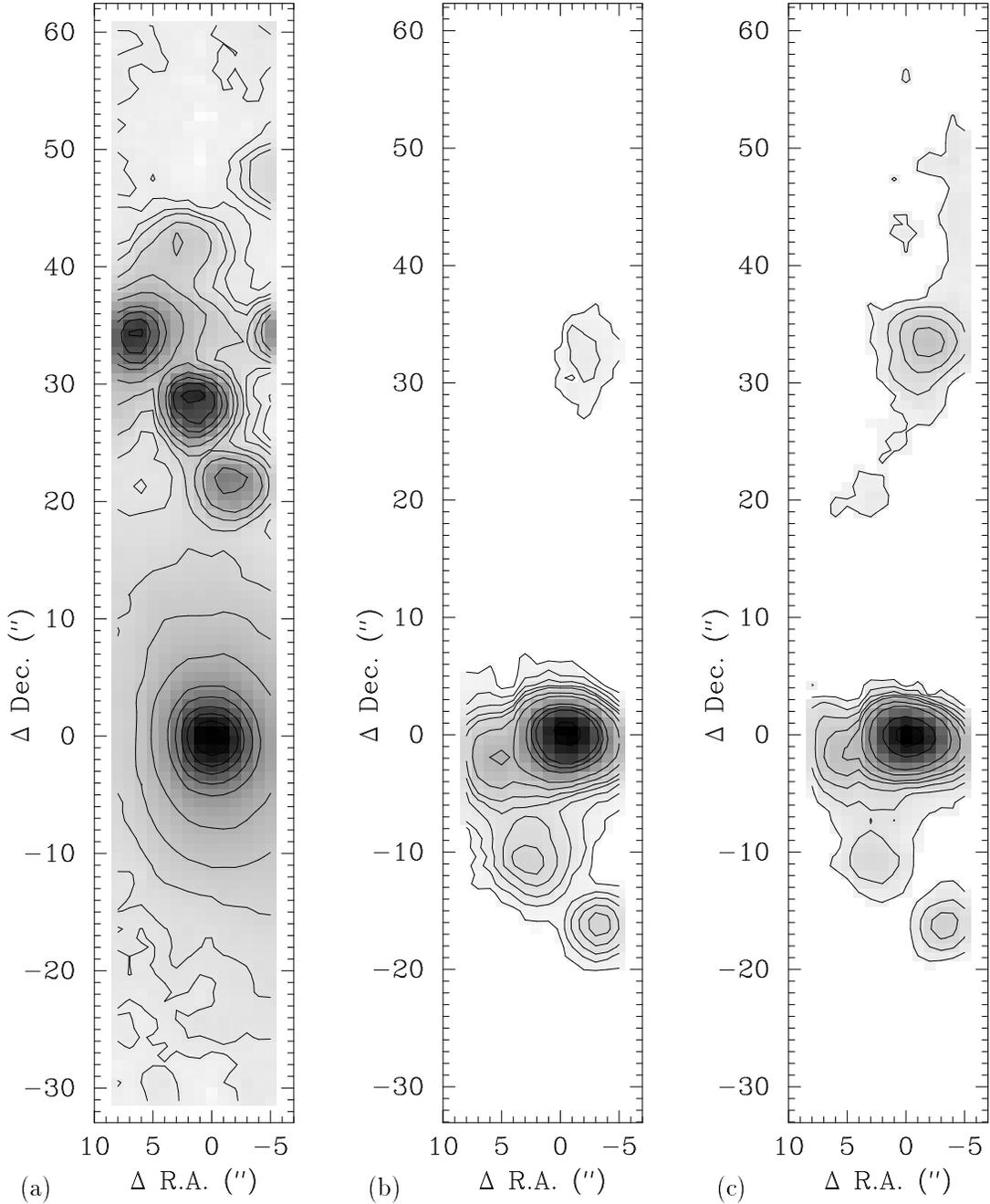,
	height=500pt}
\caption{
Images of PKS~2356$-$61 derived from the low-dispersion spectral datacube.
Positions are given in offsets of R.A. and Dec. relative to the nucleus; North
is to the top and East is to the left in all images. $(a)$ ``$V$-band''
continuum image; $(b)$ [OIII]$\lambda$5007 image; $(c)$ H$\alpha$ image. In all
images the slit is oriented north-south and pixel sizes are
$1\arcsec \times 0\farcs77$ (corresponding to a slit width of 1\arcsec, with a
spatial scale along the slit of 0\farcs77/pixel). Pixels along the slit have
not been rebinned into 1\arcsec\ increments, in order to avoid the introduction
of artifacts due to non-uniform resampling of flux. Instead, the aspect ratio
of pixels is plotted in such as way that the spatial scale in arcseconds is the
same along each axis.
}
\label{fig_img_cont}
\end{minipage}
\end{figure*}

We used the low-dispersion datacube to investigate the spatial distribution
of the stellar continuum and emission-line gas in PKS~2356$-$61. The datacube
was smoothed in R.A. and Dec. using a ``boxcar'' filter ($3 \times 3$ pixels),
in order to minimise effects due to possible changes in seeing throughout the
night. A continuum image, shown in Figure~\ref{fig_img_cont}$a$, was produced
by summing all channels with observed wavelengths in the range
$5591 - 6685$\AA\ (\ie rest wavelengths of $5100 - 6100$\AA, corresponding
approximately to $V$-band), taking care to avoid the 5577\AA\ night sky line as
well as [OIII]$\lambda$5007 and other strong emission lines. We fitted
elliptical isophotes to the host galaxy in this image using the IRAF/STSDAS
package ``ellipse''
(Jedrzejewski 1987),
\nocite{Jedrzejewski.1987.MNRAS.226.747}
%
%
and compared the results with ellipses fitted on a broad-band CCD image of the
object. In neither case did we find any significant residual continuum features
co-incident with the emission-line region; instead, the stellar continuum
follows a well-behaved elliptical light distribution. Furthermore, the radial
distribution of surface brightness, ellipticity {\it and} position angle as
derived from the spectroscopic datacube showed excellent agreement with those
obtained from the CCD image, confirming that the datacube is capable of
yielding spectrophotometric results as reliable as those obtained using
conventional imaging photometric techniques.

Some faint, extended continuum emission appears to be associated with the
objects toward the north of the source; this is also evident on the broad-band
CCD image. Spectroscopically, this structure is associated with H$\alpha$
emission at very similar systemic velocities to the gas in the central source;
it is therefore likely that this material is dynamically associated with the
central galaxy. The brighter objects, however, are foreground stars, as
confirmed by examination of their summed low-dispersion spectra.

``Narrow-band'' images of the emission-line gas in [OIII]$\lambda$5007 and
H$\alpha$ are presented in Figures~\ref{fig_img_cont}$b$,$c$. These were
obtained by parameterizing the line profiles in terms of Gaussian components.
All pixels containing detectable line emission were fitted using the
``specfit'' routine within IRAF, yielding values and error estimates for the
fluxes, velocities and widths of the lines [OIII]$\lambda$5007, H$\beta$,
H$\alpha$, [NII]$\lambda\lambda$6548,6583 and [SII]$\lambda\lambda$6716,6731.
The spectral resolution is $\sim 4$\AA, sufficiently high to separate the
H$\alpha$ and [NII] profiles but also low enough that each line profile can be
adequately described by a single Gaussian. The resulting images in
Figures~\ref{fig_img_cont}$b$,$c$ represent all pixels with flux values greater
than the corresponding formal \hbox{3-$\sigma$} limits obtained from the fits.

The emission-line region associated with the host galaxy may be described
generally in terms of two components, a bright elongated {\it central region}
and a fainter {\it tail} extending up to $\sim 15 - 20\arcsec$ from the
nucleus. The tail itself consists of three distinct clouds which we designate
{\it A}, {\it B} and {\it C}, in order of increasing distance from the central
region. No association with the radio axis (at PA~$\sim 135$\arcdeg) is
evident, either in the central region or in the tail. However, it is
interesting to note that the major axis of the central region is approximately
perpendicular to the orientation of the host galaxy optical continuum
distribution.

A large H$\alpha$-emitting region is evident to the north of the host galaxy,
displaying relatively little [OIII] $\lambda$5007 emission. This region
corresponds to the faint, diffuse emission apparent in the continuum image. A
faint extension can be seen on either side of its central component; the total
projected length of these extensions is $\sim 55\,$kpc. The systemic velocity
of this object is within $\sim 200\kms$ of that of the central galaxy, and this
is considerably smaller than expected if two cluster objects were randomly
superimposed (in which case the velocity difference should be comparable to the
projected cluster velocity dispersion, \ie $\gtrsim 1000\kms$). Thus, it is
possible that the H$\alpha$ object corresponds to a gas-rich galaxy that is
interacting with the host galaxy of PKS~2356$-$61, in which case its extended
morphology may be accounted for by tidal disruption.

\subsection{Kinematics}

The high-dispersion spectral datacube has a velocity FWHM resolution
$\sim 70\kms$, the spectral sampling being 0.8\AA. The observations were taken
at the same slit positions and offsets as those used for the low-dispersion
datacube, but the data were obtained on a different night, so that seeing
effects could potentially introduce differences in the flux distribution
compared to the low-dispersion datacube. In order to minimize such effects, the
dataset discussed here has been smoothed in the same way as the low-dispersion
dataset, by using a ``boxcar'' filter of $3 \times 3$ pixels.

In Figure~\ref{fig_cube_avs} we present several different views of the
continuum-subtracted [OIII]$\lambda$5007 datacube. These images were created
with the AVS package%
\footnote{Copyright \copyright 1993, 1994, Advanced Visual Systems, Inc.
}%
.
The orientations in all panels are such that redshift increases towards the
right. The top left panel (Fig.~\ref{fig_cube_avs}$a$) shows the flux
distribution of the gas as viewed on the sky, with north toward the top and
east to the left. The cube is then presented at a number of different viewing
angles; for example, Figure~\ref{fig_cube_avs}$d$ shows the equivalent of a
longslit spectrum covering the whole emission-line region --- in this image,
north is at the top and south is at the bottom (\ie the slit direction runs
vertically along the page).

%

\addtocounter{figure}{1}

\begin{figure*}
\centering\epsfig{file=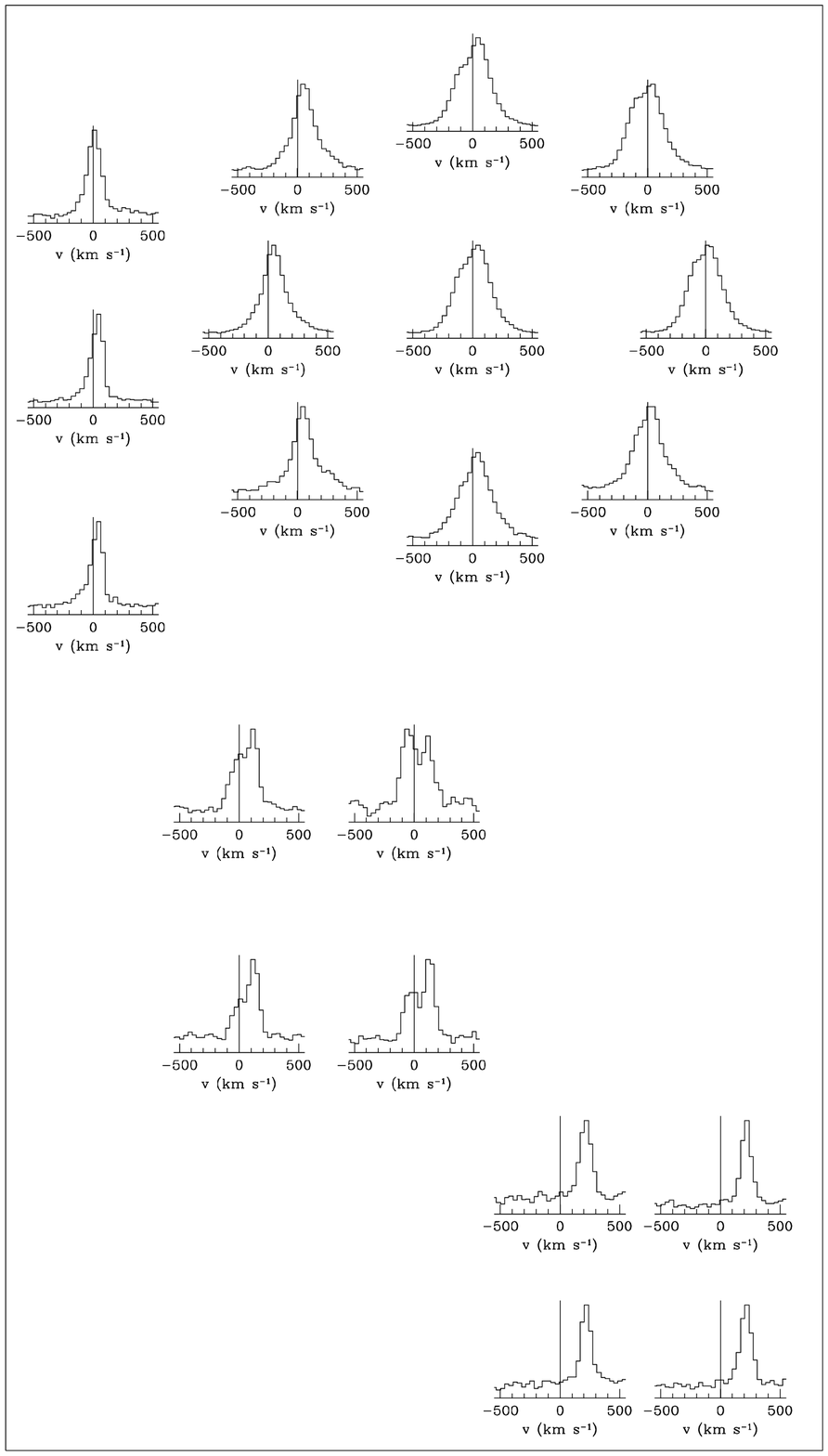,
	bbllx=109, bblly=41, bburx=500, bbury=741,
	height=498pt}
\caption{
High-dispersion [OIII]$\lambda$5007 velocity structure of the emission-line gas
in PKS~2356$-$61. Each spectrum represents the sum of a region
$\sim 3 - 4\arcsec$ across, \ie larger than the scale used for smoothing the
datacube, thus adjacent spectra are independent of one another. All velocities
are relative to the systemic velocity (mean emission-line redshift of the
entire system), which is indicated by the vertical axis. The orientation of
this image is the same as in Figure~\ref{fig_img_cont}; North is toward the top
and East is to the left.
}
\label{fig_img_vel}
\end{figure*}

A more quantitative representation of the velocity structure is shown in
Figure~\ref{fig_img_vel}; a comparison of these data with
Figure~\ref{fig_cube_avs} reveals that the three clouds comprising the tail are
spatially {\em and} kinematically distinct, having different central velocities
as well as different internal velocity fields. However, the velocity widths of
individual components in these clouds are almost all the same, ranging from 120
to 150$\kms$. Here we describe the general characteristics of the tail clouds.

Cloud {\it C} has a higher redshift than any of the other gas components in the
EELR. It contains only one velocity component, with an almost constant velocity
width. The velocity centroids also show little or no systematic change across
the entire extent of this cloud.

Cloud {\it B} is at a lower redshift and is comprised of two kinematically
distinct components, {\it B1} and {\it B2}, at lower and higher redshifts
respectively. These components both have a very similar velocity width to the
gas in {\it C}, and the width also appears to remain constant across their
entire extent. Furthermore, they overlap almost entirely on the sky and their
surface brightnesses are very similar. Component {\it B1} shows no systematic
change in velocity centroid but {\it B2} displays a smooth velocity gradient
across its entire extent, being more redshifted in the direction of {\it C}. A
faint connection may be seen between {\it B2} and {\it C}, but this is possibly
a result of the smoothing since the unsmoothed dataset contains a set of pixels
between the two clouds where there is no detectable [OIII]$\lambda$5007
emission. The velocity gradient is not a result of the smoothing since it
extends over many smoothing scales.

Cloud {\it A} contains a bright component with a width similar to that of
cloud {\it C} but it also contains a fainter, broader, blue-shifted component.
The bright component shows no systematic change in its velocity centroid, which
is intermediate between those of the two components in {\it B}. Its velocity
centroid is also very close to the value of the systemic velocity. The faint
broad blue extension in {\it A} is characteristic of a ``blue wing'' although
given its low S/N, it could also be a single separate component, as is the case
for {\it B}. The faint component is considerably broader, having approximately
the same width as the total difference between the two components in {\it B}.
A significant spatial overlap is apparent between clouds {\it A} and {\it B},
even in the unsmoothed datacube.

The central component is kinematically distinct from the clouds in the tail. It
is dominated by a broad component, with FWHM values $\sim 300 - 400\kms$, and
also has a much higher surface brightness than any of the tail regions. This
region also contains asymmetric extensions to the main component, possibly
indicating narrow components similar to those found in the tail clouds; these
could also be due to the presence of blue and red wings.

\subsection{Emission-Line Ratios}

We next consider the physical conditions and excitation state of the gas, using
the spatial variation of a number of emission-line ratios across the EELR as
derived from the low-dispersion datacube. In this sub-section we present
several emission-line ratio maps, followed by composite spectra for a number of
sub-regions within the EELR.

\subsubsection{Line-Ratio Images}

In Figure~\ref{fig_img_line} we present images of the spatial variation of the
line ratios [OIII]$\lambda$5007/H$\beta$, [NII]$\lambda$6583{\slash}H$\alpha$,
[SII]$\lambda\lambda$6716+6731/H$\alpha$, [SII]$\lambda\lambda$6716/6731, and
H$\alpha$/H$\beta$.

%

\addtocounter{figure}{1}

An important consideration in quantifying the Balmer line fluxes is the degree
of Balmer absorption present in the underlying stellar continuum. This has been
addressed in the literature using a variety of techniques
(\eg Koski \& Osterbrock 1976; Filippenko \& Sargent 1988;
Storchi-Bergmann et~al. 1992),
\nocite{Koski.1976.ApJ.203.L49,
	Filippenko.1988.ApJ.324.134,
	StorchiBergmann.1992.ApJ.396.45}
%
%
generally involving an independent determination of the stellar continuum.
However, in PKS~2356$-$61 there are no strong emission-line free regions of
continuum, and the continuum is also too faint to allow an unambiguous
determination of the intrinsic degree of absorption, for example by using
synthetic stellar spectra.

Instead, the significance of stellar absorption may be determined through a
comparison of the variation of the ratio H$\alpha$/H$\beta$ and the stellar
continuum. It can be seen in Figure~\ref{fig_img_line}$e$ that a strong peak in
this ratio is present at a location immediately to the east of the nucleus,
increasing by almost a factor of 2 to values in the range $5-6$. Since the
stellar continuum is relatively weak, a drastic change in stellar population
would be required to account for this effect by Balmer line absorption;
however, no such change is observed in the host galaxy continuum spectra. Thus
we infer that the variations in H$\alpha$/H$\beta$ across the central EELR are
indicative of intrinsic variations in the amount of extinction, and not the
result of underlying absorption. In the tail, however, there is a possibility
of significant underlying absorption in the continuum, which we discuss further
when presenting the one-dimensional low-dispersion summed spectra.

Examination of the line-ratio images reveals the presence of a ``band'' of
excitation across the nucleus, extending towards the south-west and north-east.
This feature is particularly apparent on the images of
[OIII]$\lambda$5007/H$\beta$ and [NII]$\lambda$6583/H$\alpha$, displaying
typical line ratio values of \hbox{[OIII]$\lambda$5007/H$\beta$~$\sim 10$},
[NII]$\lambda$6583/H$\alpha$~$\gtrsim 1$ and
[SII]$\lambda\lambda$6716+6731/H$\alpha$~$\sim 0.9$. Upon moving away from the
nucleus, two clear trends in excitation are evident:
(1)~towards the south-east and north-west, \ie in a direction perpendicular to
the band, the [OIII]$\lambda$5007/H$\beta$ ratio increases to around
\hbox{$15 - 20$} while [NII]$\lambda$6583/H$\alpha$ decreases to
$\sim 0.5 - 0.6$, and [SII]$\lambda\lambda$6716+6731/H$\alpha$ also decreases
to \hbox{$\sim 0.5$};
(2)~along the band, towards the south-west and north-east, the behaviour is
somewhat inverted, with [OIII]$\lambda$5007/H$\beta$ remaining at a value
$\sim 10$ and [NII]$\lambda$6583/H$\alpha$ increasing to values $\gtrsim 1.2$.
The band also coincides with the region of high reddening inferred from the
H$\alpha$/H$\beta$ ratio, suggesting that this region may correspond to a ring
or disk of dust-rich material. The one-dimensional composite spectra discussed
in the following sub-section allow a more detailed investigation of the
physical properties of various regions in the EELR.

\subsubsection{Composite Spectra}

\begin{figure*}
\begin{minipage}{\linewidth}
\centering\epsfig{file=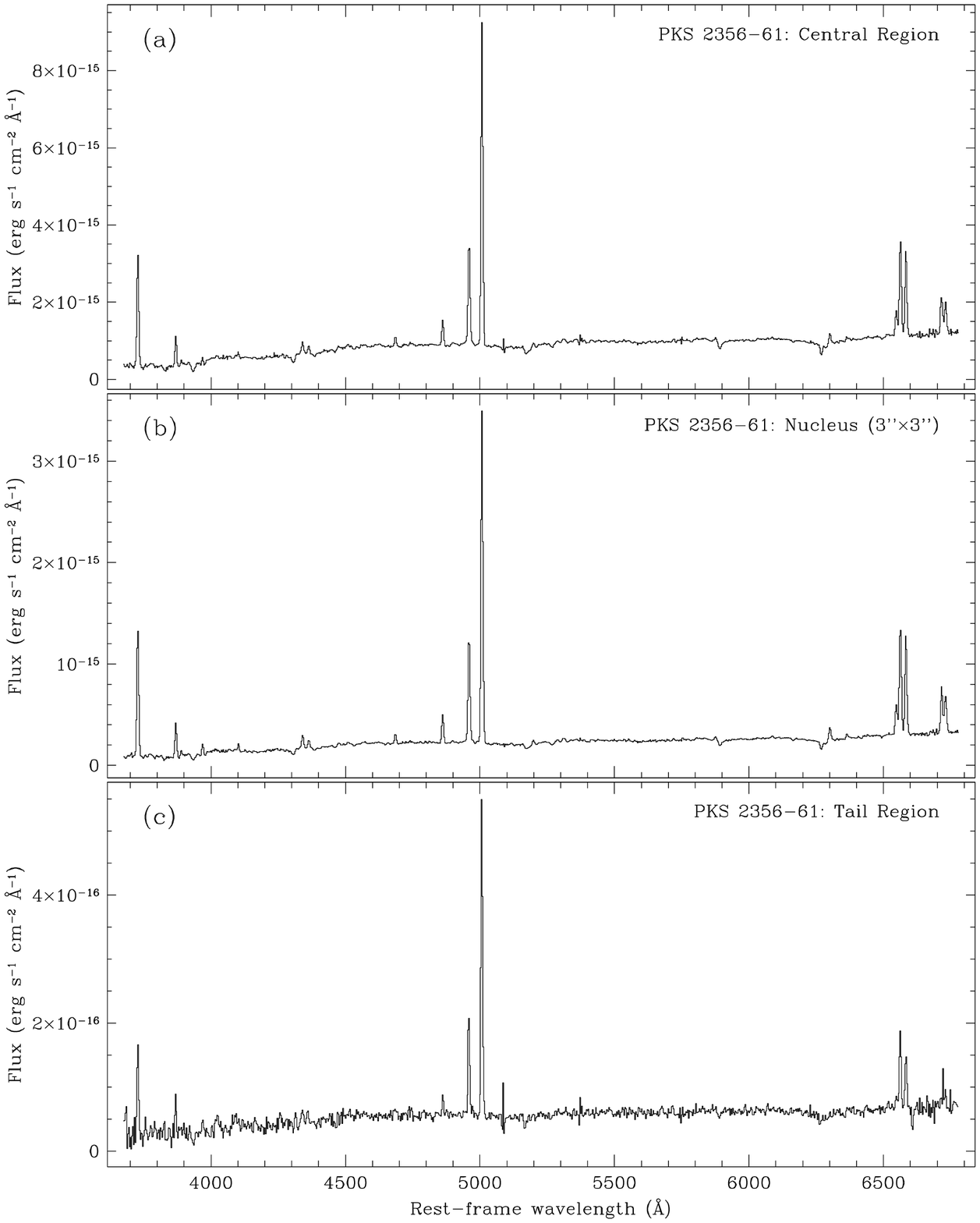,
 	height=500pt}
\end{minipage}
\caption{
Low-dispersion spectra (rest wavelength) for: $(a)$ the central region,
$(b)$ the nuclear region ($3\arcsec \times 3\arcsec$), and $(c)$ the entire
tail region. These spectra were produced by summing all the individual spectra
corresponding to pixels that have a detectable [OIII] flux in the emission-line
image, and thus represent high S/N integrated spectra for each region.
}
\label{fig_spec_1}
\end{figure*}

Low-dispersion composite spectra are presented in Figure~\ref{fig_spec_1} for
the entire central emission-line region, the nucleus
($3\arcsec \times 3\arcsec$), and the summed emission from the three tail
regions {\it A}, {\it B} and {\it C}. The wavelength scales of the spectra have
been corrected to rest wavelength, using the mean redshift of the brightest
emission lines.

\begin{figure*}
\begin{minipage}{\linewidth}
\centering\epsfig{file=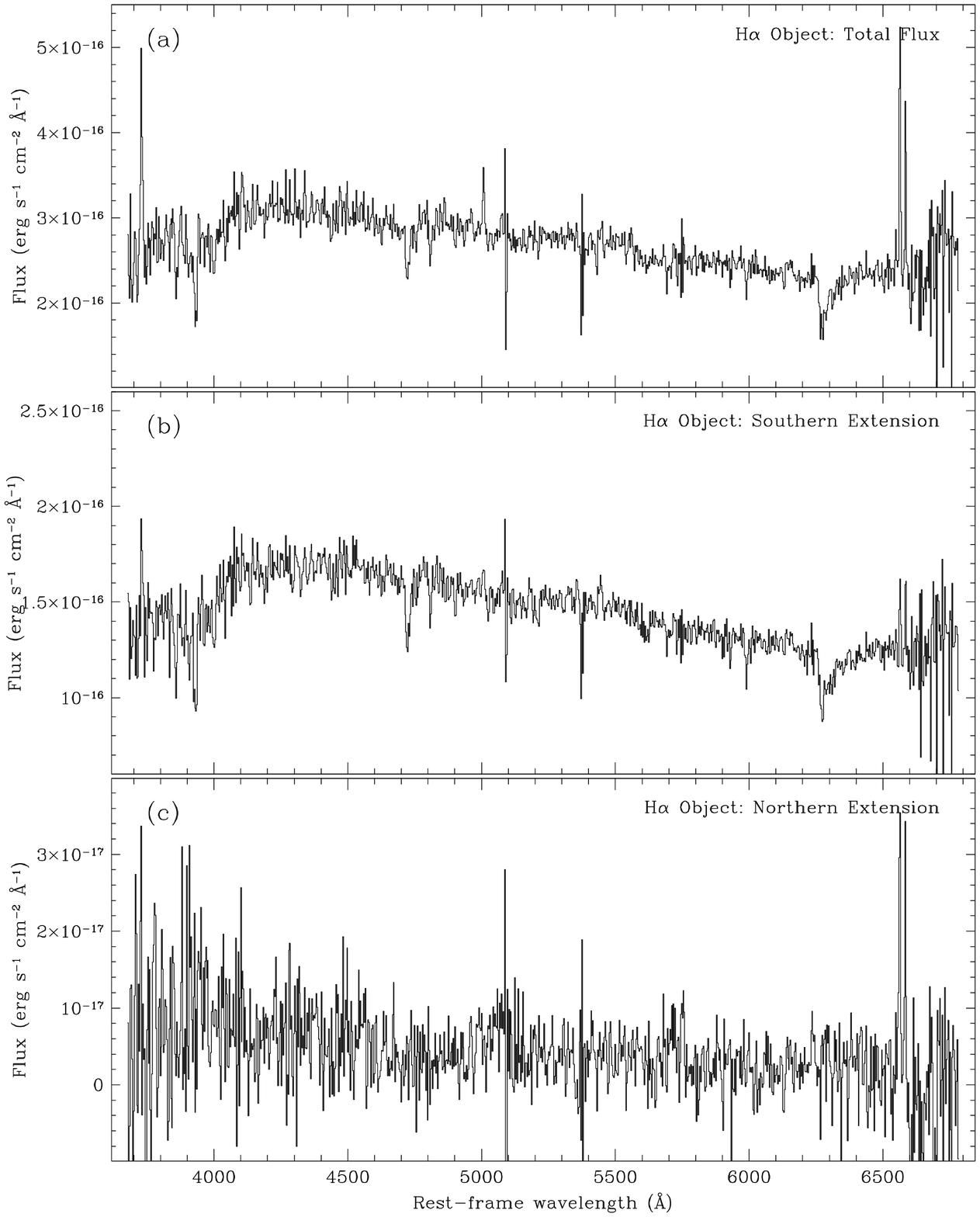,
	height=500pt}
\end{minipage}
\caption{
Low-dispersion spectra (rest wavelength) for the faint H$\alpha$ object to the
north of PKS 2356$-$61. These spectra were produced in the same way as those
presented in Figure~\ref{fig_spec_1}, by summing all the individual spectra
corresponding to pixels with a detectable H$\alpha$ flux. The top
panel (a) shows the total summed spectrum for all pixels covering the entire
H$\alpha$ object, including the northern and southern extensions and its
central region. The second and third panels (b) and (c) display the spectra for
the southern and northern extensions, and are shown here in order to afford a
comparison between the two extensions and the total spectrum, the differences
possibly being a result of different stellar populations in each region.
}
\label{fig_spec_2}
\end{figure*}

In Figure~\ref{fig_spec_2}, we present the summed spectra for the H$\alpha$
object to the north of PKS~2356$-$61. Its spectrum is typical of a star-forming
region, with a very young stellar population and strong Balmer and
[OII]$\lambda$3727 emission, with very little [OIII]$\lambda$5007. It is
interesting to note that the two extensions both show some emission-line
features, although their stellar continua are very different. In particular,
the southern extension has a spectrum very similar to the nucleus of the
object, while the northern extension has a much larger Balmer line equivalent
width, displaying very little continuum emission. This may indicate that the
northern extension consists almost exclusively of star-forming regions, while
the stronger continuum in the nucleus and southern extension reveals the
additional presence of somewhat older stars; a more definitive interpretation
is precluded by the relatively low S/N.

\begin{deluxetable}{lrlrlrlrlrl}
\tablewidth{0pt}
\tablecaption{\label{tab_fluxes}
	Line fluxes and errors for the low-dispersion spectra of PKS 2356$-$61.}
\tablehead{
Emission		&\multicolumn{2}{c}{Centre}	&\multicolumn{2}{c}{Tail (Tot.)}	&\multicolumn{2}{c}{Tail ({\it A})}	&\multicolumn{2}{c}{Tail ({\it B})}	& \multicolumn{2}{c}{Tail ({\it C})}	\\
\cline{2-3}\cline{4-5}\cline{6-7}\cline{8-9}\cline{10-11}
line			& Flux	& 1-$\sigma$	& Flux	& 1-$\sigma$	& Flux	& 1-$\sigma$	& Flux	& 1-$\sigma$	& Flux	& 1-$\sigma$	}
\startdata
[OII]$\lambda$3727	& 226.0	& 3.4		& 10.9	& 0.6		& 5.14	& 0.87		& 3.26	& 0.82		& 1.03	& 0.65		\nl
[NeIII]$\lambda$3869	& 54.3	& 2.1		& 4.53	& 0.51		& 2.20	& 0.64		& \nodata & \nodata	& \nodata & \nodata	\nl
H$\gamma$		& 35.8	& 1.7		& 2.45	& 0.45		& 0.47	& 0.36		& \nodata & \nodata	& \nodata & \nodata	\nl
[OIII]$\lambda$4363	& 20.7	& 1.6		& 1.88	& 0.40		& \nodata & \nodata	& \nodata & \nodata	& \nodata & \nodata	\nl
HeII$\lambda$4686	& 19.4	& 1.2		& 1.84	& 0.33		& \nodata & \nodata	& \nodata & \nodata	& \nodata & \nodata	\nl
H$\beta$		& 53.0	& 1.2		& 2.35	& 0.43		& 1.33	& 0.26		& 0.60	& 0.29		& 0.47	& 0.21		\nl
[OIII]$\lambda$5007	& 662.0	& 1.8		& 38.3	& 0.7		& 19.4	& 0.5		& 11.0	& 0.4		& 4.48	& 0.28		\nl
[NI]$\lambda$5200	& 9.95	& 1.06		& 0.67	& 0.41		& 0.35	& 0.23		& \nodata & \nodata	& \nodata & \nodata	\nl
[OI]$\lambda$6300	& 32.7	& 1.2		& 0.60	& 0.35		& 0.33	& 0.21		& \nodata & \nodata	& \nodata & \nodata	\nl
H$\alpha$		& 229.0	& 1.4		& 10.1	& 0.6		& 4.98	& 0.26		& 2.54	& 0.28		& 1.68	& 0.23		\nl
[NII]$\lambda$6583	& 203.0	& 3.7		& 7.12	& 0.57		& 3.02	& 0.27		& 2.18	& 0.30		& 1.11	& 0.22		\nl
[SII]$\lambda$6716	& 95.0	& 2.2		& 1.27	& 0.63		& 0.93	& 0.42		& \nodata & \nodata	& \nodata & \nodata	\nl
[SII]$\lambda$6731	& 70.3	& 1.7		& 1.67	& 0.58		& 0.77	& 0.37		& \nodata & \nodata	& \nodata & \nodata	\nl
\enddata
\tablecomments{
All values are in units of $1 \times 10^{-16}\,$erg$\,$s$^{-1}\,$cm$^{-2}$.
}
\end{deluxetable}

\begin{deluxetable}{lrrrrr}
\tablewidth{0pt}
\tablecaption{\label{tab_ratios}
	De-reddened line ratios for PKS 2356$-$61.}
\tablehead{
Line Ratio				&\multicolumn{1}{c}{Centre}	&\multicolumn{1}{c}{Tail (Tot.)}&\multicolumn{1}{c}{Tail ({\it A})}	&\multicolumn{1}{c}{Tail ({\it B})}	& \multicolumn{1}{c}{Tail ({\it C})}
}
\startdata
$C$					& 1.169	~			& 1.189	~			& 0.761	~			& 1.125	~			& 0.647	~			\nl
\tablevspace{3pt}
HeII$\lambda$4686/H$\beta$		& 0.39	$_{0.36}^{0.42}$	& 0.84	 $_{0.62}^{1.13}$	& \nodata\hfill			& \nodata\hfill			& \nodata\hfill			\nl
[OIII]$\lambda$5007/H$\beta$		& 11.9	$_{11.6}^{12.2}$	& 16	~$_{14}^{19}$		& 14	~$_{12}^{17}$		& 18	$_{12}^{30}$		& 9.5	$_{~7.1}^{13.3}$	\nl
[OIII]$\lambda\lambda$5007+4959/4363	& 34.5	$_{31.9}^{37.5}$	& 22	~$_{19}^{27}$		& \nodata\hfill			& \nodata\hfill			& \nodata\hfill			\nl
[OIII]$\lambda$5007/[OI]$\lambda$6300	& 27.4	$_{26.4}^{28.6}$	& 87	 $_{~57}^{179}$		& 72	 $_{~47}^{146}$		& \nodata\hfill			& \nodata\hfill			\nl
[OIII]$\lambda$5007/[OII]$\lambda$3727	& 1.90	$_{1.87}^{1.94}$	& 2.3	~$_{2.2}^{2.4}$		& 2.8	~$_{2.5}^{3.3}$		& 2.2	$_{1.8}^{2.9}$		& 3.4	$_{2.4}^{5.9}$		\nl
[OIII]$\lambda$5007/[NII]$\lambda$6583	& 4.69	$_{4.59}^{4.79}$	& 7.9	~$_{7.3}^{8.5}$		& 8.2	~$_{7.5}^{8.9}$		& 7.2	$_{6.3}^{8.3}$		& 4.9	$_{4.3}^{5.7}$		\nl
[OIII]$\lambda$5007/H$\alpha$		& 4.15	$_{4.12}^{4.19}$	& 5.6	~$_{5.3}^{6.0}$		& 4.9	~$_{4.7}^{5.2}$		& 6.2	$_{5.5}^{6.9}$		& 3.3	$_{3.0}^{3.6}$		\nl
[OI]$\lambda$6300/H$\alpha$		& 0.15	$_{0.14}^{0.16}$	& 0.06	 $_{0.03}^{0.10}$	& 0.07	 $_{0.03}^{0.11}$	& \nodata\hfill			& \nodata\hfill			\nl
[SII]$\lambda\lambda$6716+6731/H$\alpha$ & 0.70	$_{0.68}^{0.72}$	& 0.28	 $_{0.21}^{0.37}$	& 0.34	 $_{0.20}^{0.48}$	& \nodata\hfill			& \nodata\hfill			\nl
[NII]$\lambda$6583/H$\alpha$		& 0.89	$_{0.87}^{0.91}$	& 0.71	 $_{0.63}^{0.79}$	& 0.60	 $_{0.54}^{0.67}$	& 0.86	 $_{0.70}^{1.05}$	& 0.66	 $_{0.56}^{0.78}$	\nl
[SII]$\lambda\lambda$6716/6731		& 1.35	$_{1.29}^{1.42}$	& 0.76	 $_{0.36}^{1.46}$	& 1.2	~$_{0.6}^{2.7}$		& \nodata\hfill			& \nodata\hfill			\nl
\enddata
\tablecomments{
The ratio upper and lower limits are derived from the \hbox{1-$\sigma$} flux
errors.
}
\end{deluxetable}

Line fluxes were measured using the IRAF routine ``specfit''; since these
spectra are of low dispersion, each profile can be adequately described by a
single Gaussian component. In clouds {\it B} and {\it C}, the [SII] lines were
not fitted due to the noise contribution from skylines at 7300\AA.
Table~\ref{tab_fluxes} lists the measured fluxes and \hbox{1-$\sigma$} rms
errors for all the lines fitted. The errors are the formal \hbox{1-$\sigma$}
errors obtained from the fits, but they do not represent other effects such as
sky background subtraction uncertainties. Thus their primary use is simply to
give an indication of the {\em relative} accuracy of the various measured
fluxes, particularly relevant to the fainter regions.

In Table~\ref{tab_ratios} we present a number of physically important line
ratios, corrected for reddening effects using several standard assumptions.
The intrinsic ratio $I_{\lambda_1}/I_{\lambda_2}$ of the intensities of two
lines at wavelengths $\lambda_1$,$\lambda_2$, given an observed ratio
$I_{\lambda_{1,0}}/I_{\lambda_{2,0}}$ and an extinction function $f(\lambda)$
(\eg Savage \& Mathis 1979),
\nocite{Savage.1979.ARAA.17.73}
%
%
is calculated following
Osterbrock (1989):
\nocite{Osterbrock.1989.AGN2}
%
%
%
\begin{displaymath}
	{I_{\lambda_1} \over I_{\lambda_2}} =
		{I_{\lambda_{1,0}} \over I_{\lambda_{2,0}}} \,
		e^{C[f(\lambda_1) - f(\lambda_2)]}
\end{displaymath}
where the optical depth at $\lambda$ is $\tau_{\lambda} = C\,f(\lambda)$,
parameterized by the {\it coefficient of extinction} $C$.
Extinction due to our galaxy is assumed negligible as a result of the
relatively high galactic latitude of the source ($-55\arcdeg\,$S), thus all
reddening is considered to be internal to the source. Case~B recombination is
assumed, with the canonical H$\alpha$/H$\beta$ ratio of 2.87. Although some
authors claim evidence of a slightly higher intrinsic ratio of $\sim 3.0 - 3.1$
when a hard ionizing spectrum is assumed
(Halpern \& Steiner 1983; Ferland \& Osterbrock 1985; Tsvetanov \& Yancoulova
1989),
\nocite{Halpern.1983.ApJ.269.L37,
	Ferland.1985.ApJ.289.105,
	Tsvetanov.1989.MNRAS.237.707}
%
%
we find that this difference has a negligible effect on the present results.
Underlying Balmer-line absorption is also taken to be negligible, as we
discussed previously based on the H$\alpha$/H$\beta$ ratio and the continuum
flux. Finally, it is assumed that reddening for all emission-lines can be
described by the observed Balmer decrement, and that the electron temperature
$T_e$ derived from the [OIII]$\lambda\lambda$5007/4363 ratio can be applied to
the Balmer-line emitting region. In practice, different ionic species are
probably not precisely co-incident; however, dust variations within the EELR
are likely to be less significant than foreground reddening, and furthermore
the reddening values observed in PKS~2356$-$61 are sufficiently small that such
differences are likely to be undetectable.

\subsection{Physical Conditions in the Gas\label{sec3.4_PhysCond}}

\subsubsection{Electron Density and Temperature}

We use the line flux ratios of [SII]$\lambda\lambda$6716/6731 and
[OIII]$\lambda\lambda$5007/4363 to constrain the electron density $n_e$ and
temperature $T_e$ of the gas. For the central region, we find
$T_e = 2.2 \pm 0.1 \times 10^4\,$K and constrain
$n_e \sim 22 - 140\,$cm$^{-3}$, with a most probable value of
$n_e = 80\,$cm$^{-3}$. The three tail regions are treated as a single region,
distinct from the central region, since the differences between the tail and
central regions are generally much greater than within the tail, and moreover
S/N limitations allow only one value to be obtained for the entire tail region.
For the tail region, we measure $T_e = 3.2 \pm 0.6 \times 10^4\,$K and
constrain $n_e$ to lie in the range $\sim 10 - 10^4\,$cm$^{-3}$ (with a most
probable value of $n_e \sim 100\,$cm$^{-3}$). The upper and lower bounds given
for each quantity are derived from the errors in the measured ratios in
Table~\ref{tab_ratios}. We note that the values derived from these ratios may
{\it not} be representative of the line-emitting gas as a whole, as its
inferred ionization structure is model-dependent. Therefore, the derived
densities/temperatures are indicative of mean values, possibly integrated over
several zones of different ionization and density.

The [OIII] temperature is comparatively large for both the central region and
the tail. It is probable that the errors on the measurement for the tail region
are larger than the formal \hbox{1-$\sigma$} errors yielded by the flux
fitting, although we find that the maximum possible uncertainty introduced by
the underlying continuum results in a lower limit of
$T_e \sim 2.1 \times 10^4\,$K for the tail region. It should be noted that
substantial [OIII]$\lambda$4363 emission is observed across the {\it entire}
central region ($\sim 12\arcsec \times 9\arcsec$); less than 30\% of the flux
is produced by the inner $3\arcsec \times 3\arcsec$ centred on the nucleus.
Thus, the [OIII]$\lambda\lambda$4363/5007 ratio indicates an electron
temperature $T_e \gtrsim 2 \times 10^4\,$K across an {\it extended} region of
$\gtrsim 20\,$kpc; we discuss this interesting result in more detail later in
this paper.

\subsubsection{Ionization rate}

We derive the ionization rate in the gas, $N_{\rm ion}$, using the observed
H$\alpha$ luminosity in various regions of the EELR, and following standard
assumptions such as Case~B ionization equilibrium conditions
(\eg Osterbrock 1989).
\nocite{Osterbrock.1989.AGN2}
%
%
The various values of $N_{\rm ion}$ are presented in Table~\ref{tab_Nion}
(calculated from the H$\alpha$ fluxes in Table~\ref{tab_fluxes}). The volume of
each region is derived from its projected spatial extent by assuming an
approximately ellipsoidal three-dimensional geometry, with a line-of-sight
depth equal to the mean of its minimum and maximum extent on the sky (in fact,
regions {\it B} and {\it C} are represented by spherical volumes of diameter
$d$). We also tabulate the derived quantity $n_e f^{1/2}$, where $f$ represents
the volume filling factor of line-emitting gas. It should be noted that this is
only an approximate indication of the electron density, applicable to material
in ionization equilibrium; the occurrence of shocks can produce post-shock
density enhancements two to three orders of magnitude higher than the precursor
gas.

\begin{deluxetable}{lcccc}
\tablewidth{0pt}
\tablecaption{\label{tab_Nion}
	Ionization rates in the EELR.}
\tablehead{
		& Extent		& Volume	& $N_{\rm ion}$		& $n_e f^{1/2}$	\\
		& (arcsec)		& (kpc$^3$)	& (s$^{-1}$)		& (cm$^{-3}$)
}
\startdata
Centre		& $10 \times 8$		& 2425		& $3.14 \times 10^{53}$ & 0.176 \nl
Tail {\it A}	& $\phn6 \times 9$	& 1365		& $6.83 \times 10^{51}$	& 0.034 \nl
Tail {\it B}	& $d = 7\arcsec$	& 1155		& $3.48 \times 10^{51}$	& 0.027 \nl
Tail {\it C}	& $d = 6\arcsec$	& \phn730	& $2.30 \times 10^{51}$	& 0.028 \nl
\enddata
\end{deluxetable}

Later in this paper, we use the values of $N_{\rm ion}$ to investigate various
possible ionization mechanisms for the tail clouds, as well as the covering
fractions of the clouds to ionizing radiation, and their geometrical location
with respect to the nucleus and the plane of the sky. The limits on the
electron density from the quantity $n_e^2 f$ provide valuable additional
constraints to those derived previously from the [SII] line ratios:
if shocks are indeed a dominant ionization mechanism in the gas, it is likely
that the densities suggested by the [SII] ratios apply primarily to the
post-shock material, while the precursor densities may be closer to the lower
limits tabulated in Table~\ref{tab_Nion}. In the following section we discuss
in more detail the question of the relative contributions of precursor and
post-shock emission, through a detailed examination of observed line ratios,
fluxes and velocity widths, and a comparison with results from auto-ionizing
shock models.

\section{Emission-line Energetics and Excitation\label{EELR_sec4}}

Here we investigate the applicability of the auto-ionizing shock model to the
EELR, by examining three complementary observable quantities:
(a)~{\em velocity width} along the line-of-sight;
(b)~{\em surface brightness} for a region of particular velocity width;
(c)~{\em line ratios} for a region of particular velocity width.
If shocks are the principal ionization mechanism in the gas then these three
observables should be inter-related; this allows good constraints to be placed
upon the consistency of the model. The comparison between the shock model and
the observations is carried out in three different ways:
\newcounter{tmpii}\newlength{\tmplngth}
\begin{list}
{\arabic{tmpii}.}{\usecounter{tmpii}\setlength{\leftmargin}{0in}
\setlength{\tmplngth}{\parindent}\addtolength{\tmplngth}{\labelwidth}
\addtolength{\tmplngth}{-\labelsep}\setlength{\itemindent}{\tmplngth}}
\item
{\bf Energetic feasibility}. Two aspects are considered: (a)~The variation of
observed luminosity as a function of velocity width, for all pixels, as
compared with the relationships expected from the shock model; (b)~Can the EELR
luminosity be accounted for energetically by shocks with velocities
corresponding to the observed velocity widths?
\item
{\bf Excitation: Dependence on {\boldmath$\Delta v$}.} This comparison is made
by examining the excitation state of each pixel (quantified using a ratio of
two emission lines) as a function of its velocity width. Resulting trends are
compared with those expected on the basis of the shock model. This comparison
uses the values measured for each pixel and is therefore limited to the
brightest lines, observable across the entire EELR.
\item
{\bf Excitation: Diagnostic line ratios.} The degree of excitation can also be
quantified by plotting pairs of line ratios against each other, specifically
aimed at separating regions where different excitation mechanisms operate. In
this comparison, integrated line fluxes are obtained by summing spectra from
different regions of the EELR, in order to obtain good signal-to-noise line
fluxes for the fainter emission lines, some of which are crucial for
determination of physical properties of the gas. The location of the resulting
points, relative to the loci from shock models, yields a powerful means of
verifying (independently of the previous two comparisons) whether the observed
excitation state is consistent with the observed velocity width if the shock
model applies.
\end{list}
In these comparisons, we assume that the observed line-of-sight velocity width
$\Delta v$ represents one or more shocks with that velocity, rather than a
superposition of kinematically unrelated systems. Observationally, we define
$\Delta v$ as the largest line-of-sight velocity difference for emission in a
given pixel. For strongly blended high S/N profiles, we find that the second
moment of the profile provides a robust means of quantifying its velocity
width; otherwise if multiple peaks are clearly apparent then $\Delta v$ is
taken to be the largest velocity difference between separate components.
The resulting values of $\Delta v$ were ``de-blended'' by the velocity
resolution of 70$\kms$; however this is not a substantial correction.

\subsection{Auto-Ionizing Shock Models}

In order to investigate the viability of auto-ionizing shocks as a possible
ionization mechanism for the gas,
Koekemoer (1996)
\nocite{Koekemoer.1996.PhD}
%
%
has computed a grid of models extending across two of the primary shock
parameters: the shock velocity and the magnetic field parameter $B/n^{1/2}$,
where $B$ is the value of the magnetic field strength transverse to the shock,
and $n$ is the precursor gas density. The gas is assumed to be in the
low-density limit in all cases. The shock models were computed using the
non-equilibrium ionization code MAPPINGS~II, described in
Sutherland \& Dopita (1993),
\nocite{Sutherland.1993.ApJS.88.253}
%
%
together with the same atomic abundances. These abundances are generally the
same as those in the Solar neighbourhood; the question of possible abundance
variations is explored further in some of the subsequent sections.

In Figures~\ref{fig_fluxes_obs}~--~\ref{fig_diagnostic_3} we present results
from the shock ionization models together with the observational determinations
of emission-line fluxes and velocity widths. The model results plotted are:
pure shocks (ranging from $150 - 550\kms$), pure precursor spectra produced by
photoionization from shocks with velocities in the range $200 - 550\kms$, and a
set of loci for the combined spectra of photon-bounded precursors and shocks in
the velocity range $200 - 550\kms$. Results are also plotted for different
values of the magnetic field parameter:
$B/n^{1/2} = 0, 1, 4\,\mu$G$\,$cm$^{3/2}$.

An increase in the magnetic parameter reduces the compression factor behind the
shock, thereby increasing the effective ionization parameter in the
photoionization-recombination region and changing the line ratios accordingly.
However, the {\it ionizing} spectrum of the shock, hence also the precursor
emission, is insensitive to the value of $B/n^{1/2}$. Thus each shock velocity
produces a single photoionized precursor spectrum, regardless of different
values of the magnetic field parameter.

\subsection{Energetic Feasibility}

We first investigate whether the emission-line regions may feasibly be powered
by shocks with velocities corresponding to the observed values of $\Delta v$.
We compare the observations with the modelling results by assuming that for a
given detector pixel, the observed velocity width corresponds to one or more
shocks along the line-of-sight having {\it only} that velocity. This is
physically justified by the strong dependence of shock luminosity upon
velocity, such that the fastest shock in any region can be expected to dominate
the spectrum.

The total surface area of shocks along the line-of-sight corresponding to a
single pixel is parameterized by the quantity $f_A$, normalised to the pixel
surface area. This is necessary because the resolution scale is of order
1\,kpc, which may be substantially larger than the scale of shocks in the
gas. Therefore it is not possible to directly constrain orientation effects for
individual shocks, or the possible presence of multiple shocks of different
velocities along the line-of-sight sampled by a single pixel; the inclusion of
such effects would introduce too many unknown parameters into the problem.
Rather, the approach taken here is to examine the degree of consistency of
shock models with the observations using a minimum of free parameters.

\begin{figure*}
\begin{minipage}{\linewidth}
\centering\epsfig{file=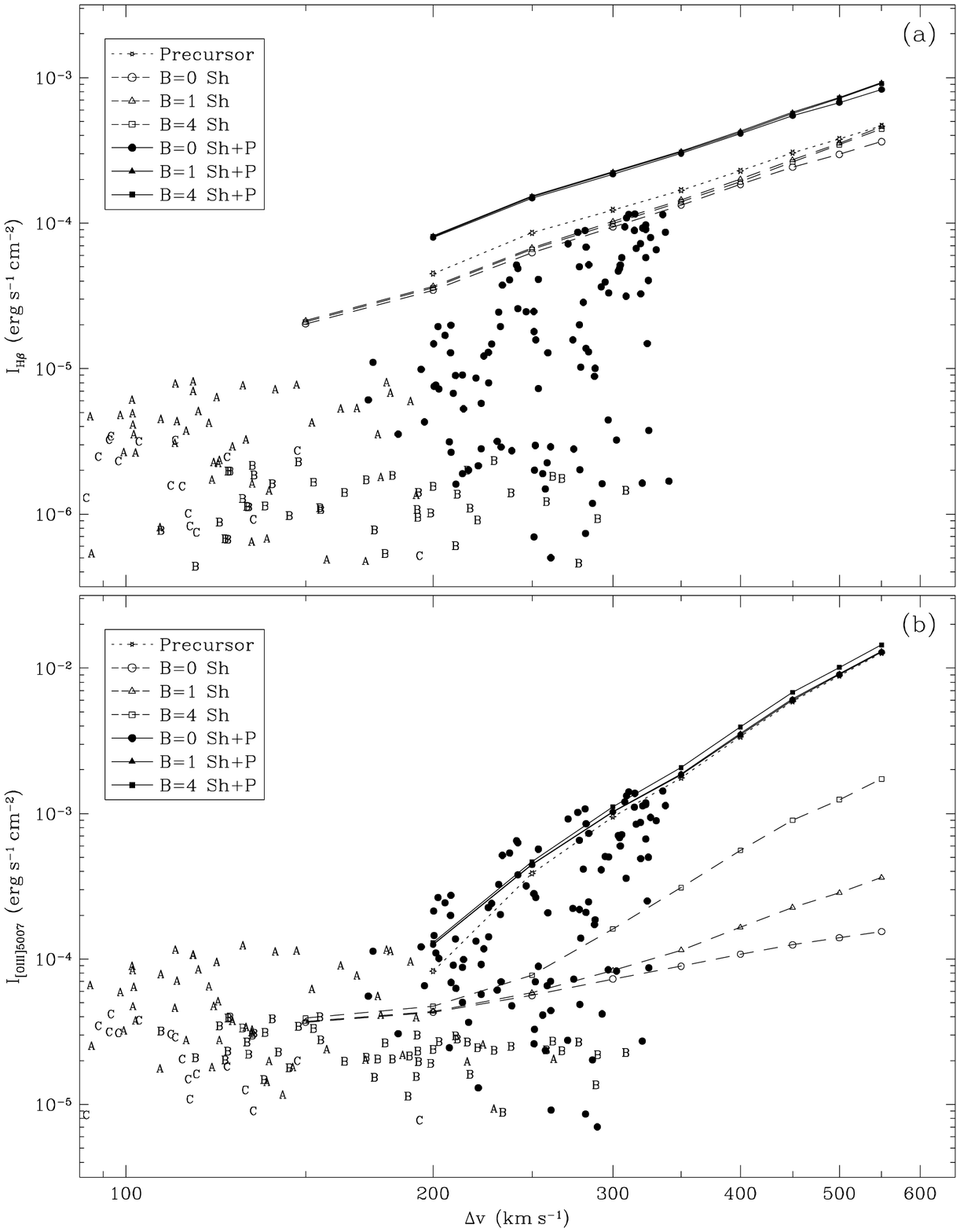,
	height=500pt}
\end{minipage}
\caption{
Variation of the intensity of $(a)$~H$\beta$ and $(b)$~[OIII]$\lambda$5007 for
each pixel in the EELR of PKS~2356$-$61, as a function of the line-of-sight
velocity width of each pixel. Points from the ``Central Region'' are indicated
by filled dots, while those from the Tail regions {\it A}, {\it B} and {\it C}
are marked accordingly. Auto-ionizing model curves are plotted for ``pure''
shocks (``Sh''), ``pure'' precursors, and combined shock and precursor emission
(``Sh+P'').  Note that the value of the $B$-parameter significantly affects
emission from the shocked region, but not from the precursor material.
}
\label{fig_fluxes_obs}
\end{figure*}

Given the above scenario, the observed flux at each pixel is converted to an
intensity, in units of erg\,s$^{-1}$ emitted per cm$^2$ of material along the
line-of-sight, thus allowing a direct comparison with the intensities predicted
by shock models. The [OIII]$\lambda$5007 and H$\beta$ intensities for all
pixels across the EELR are plotted in Figure~\ref{fig_fluxes_obs}, as a
function of the observed velocity width of each pixel. The models on the plots
represent the intensity that a pixel of a given velocity width {\it would}
have, given the above assumptions. The models are calculated for a precursor
density $n = 1\,$cm$^{-3}$ and $f_A = 1$; it should be noted that the predicted
intensity is linearly proportional to density.

The models are not intended to be a ``fit'' to the data points. Rather, the
primary result from this comparison is:
\begin{quote}
For any given point (\ie pixel) in the gas with a particular velocity width,
{\it shocks of a corresponding velocity can account for all the flux observed
in that pixel for values of $n f_A$\,$\sim$\,0.1\,$-$\,10}.
\end{quote}
There is no reason {\it a priori} to expect the shock model predictions to be
at all comparable to the observed intensities. The fact that agreement to
better than an order of magnitude is obtained for the EELR energetics, using
physically reasonable approximations, provides strong support for the argument
that shocks, if present, can readily dominate the EELR ionization energetics.

It is clear that the upper envelope of pixel intensities increases strongly as
a function of $\Delta v$, rising by an order of magnitude across a factor of
two in velocity width. In particular, {\it none} of the brightest pixels
display low velocity widths, and this cannot be explained through the presence
of any kind of observational selection effect. Indeed, observational selection
effects should be more prominent at low intensity levels: for example, weak
lines should be preferentially detected at small velocity widths. The fact that
all the low-intensity pixels display the full range of velocity widths
indicates that these observations are of sufficient S/N not to suffer from such
selection effects. The plots show that the EELR contains pixels with
(projected) shock velocities continuously ranging between $\sim 200$ and
350$\kms$, while the spread in pixel intensities of similar velocity width may
indicate different amounts of line-emitting gas.

It is also evident that for all values of $\Delta v$ above
\hbox{$\sim 200\kms$}, the observed [OIII]$\lambda$5007 flux, relative to
H$\beta$, remains about a factor of 2 higher than predicted by models involving
combined shock and precursor emission. We show in the following sub-section
that such a constant offset suggests geometrical orientation effects, in which
an inclination of the shock plane to the line-of-sight leads to an observed
velocity that is lower than the intrinsic shock velocity.

The energetics alone cannot provide further significant insights into the
physical processes along a given line-of-sight, since both the shock surface
area and the density are free parameters. In particular, it is difficult to
constrain the relative importance of emission from the shocked gas versus that
of material photoionized by the shocks. In the following sub-section we
investigate line ratio variations, which offer a powerful complementary means
of constraining the relevant physics.

\subsection{Excitation and Velocity Width}

\begin{figure*}
\begin{minipage}{\linewidth}
\centering\epsfig{file=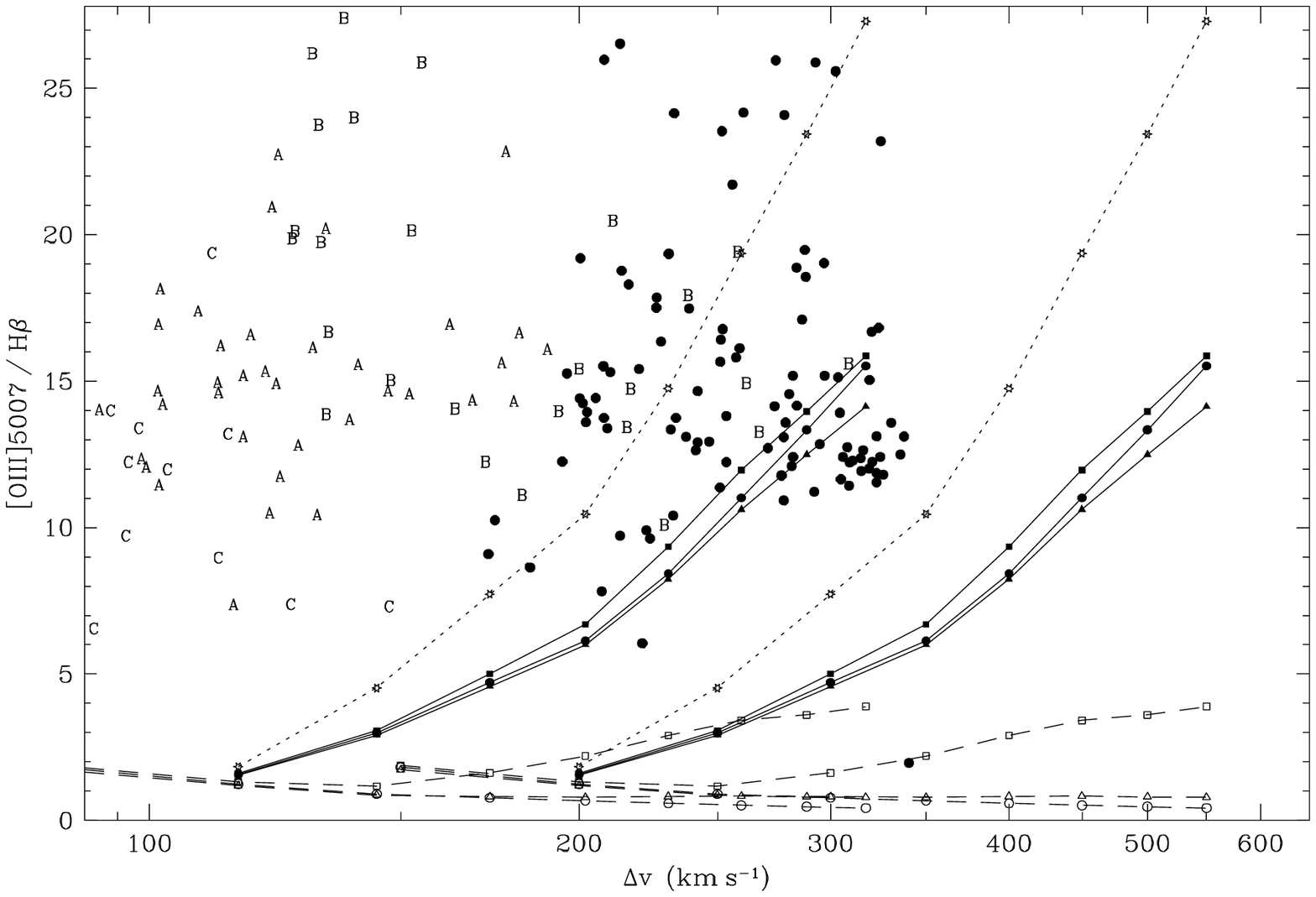,
	width=430pt}
\end{minipage}
\caption{
Variation of [OIII]$\lambda$5007/H$\beta$ as a function of observed velocity
width, for all pixels of sufficient S/N in the EELR of PKS~2356$-$61. The data
points and theoretical shock loci are represented the same way as in
Figure~\ref{fig_fluxes_obs}. Two families of shock loci are plotted (both
extending up to intrinsic shock velocities of 550$\kms$): one for a
line-of-sight inclination of 0$\arcdeg$ (\ie the shock velocity is identical to
the observed line-of-sight velocity width, extending up to 550$\kms$), while
the other corresponds to the velocity widths expected for a random distribution
of line-of-sight inclinations; (thus the observed ensemble of shock velocities
are reduced by the ``mean'' projection factor of $1/\surd3$, and only extend up
to $\sim 300\kms$).
}
\label{fig_ratios_obs}
\end{figure*}

The free parameters of shock surface area and gas density can be removed by
investigating the variation of line {\it ratios} as a function of velocity
width. In particular, a high-excitation line such as [OIII]$\lambda$5007
compared to H$\beta$ provides a good diagnostic of the relative contributions
of shock and precursor emission. In Figure~\ref{fig_ratios_obs} we present the
ratio [OIII]$\lambda$5007/H$\beta$ as a function of line-of-sight velocity
width, for all pixels with detected H$\beta$ fluxes \hbox{$\gtrsim 3$$\sigma$}.
The values of the [OIII]$\lambda$5007/H$\beta$ ratio in
Figure~\ref{fig_ratios_obs} range predominantly between 10 and 20, at all the
observed velocity widths. However, the points separate clearly into the
following two regimes, depending upon which part of the EELR they represent.

Firstly, the central region and the tail region {\it B} both display velocities
up to $\sim 350\kms$ (typically in the range $250 - 300\kms$); their observed
values of [OIII]$\lambda$5007/H$\beta$ correspond to a combination of emission
from photoionized precursor material (kinematically associated with shocks) and
cooling material behind shocks of velocities \hbox{$\sim$ 400 $-$} 550$\kms$.
These shock velocities become consistent with the observed velocity widths upon
invoking the standard projection factor of $1/\surd3$ for an ensemble of
line-of-sight shock inclination angles (thereby producing observed velocities
only $\sim 50 - 60\%$ of the intrinsic shock velocities).

Secondly, the tail regions {\it A} and {\it C} display velocity widths almost
exclusively $\lesssim 200\kms$ and predominantly in the range $100 - 150\kms$,
much too small to be kinematically associated with shocks of the velocities
required to produce the observed [OIII]$\lambda$5007/H$\beta$ ratios (unless
highly improbable geometric situations are invoked). Thus, in the context of
auto-ionizing shocks these regions correspond to gas photoionized by shocks,
but not kinematically associated with such shocks.

If the line-emitting gas in the central region is related to infalling
material, for example gas captured tidally through an interaction with a
companion galaxy, then such gas clouds are likely to virialize on significantly
shorter timescales than the overall lifetime of the source. If shocks in this
region are the result of collisions between individual clouds, then the
virialized velocity distribution must be deprojected to correspond to the
observed velocities. We use the standard deprojection factor of $1/\surd3$ for
an ensemble of clouds that possess random velocity vectors. The actual
situation is likely to be more complex: all the clouds would not be moving at
identical speeds; shock velocities would depend upon the relative trajectories
of the clouds involved; and the cloud velocity vectors are not likely to be
completely randomized. However, it is clear that agreement with the
observations is substantially improved if the observed velocities correspond to
$\sim 50 - 60\%$ of their intrinsic values, not unreasonable for an ensemble of
clouds in the process of virialization.

\subsection{Diagnostic Line Ratios}

The final comparison between observed emission-line properties and those
predicted from shock models, involves the use of ``diagnostic'' line-ratio
plots
(Baldwin et~al. 1981; Veilleux \& Osterbrock 1987; Dopita \& Sutherland 1996).
\nocite{Baldwin.1981.PASP.93.5,
	Veilleux.1987.ApJS.63.295,
	Dopita.1996.ApJS.102.161}
%
%
Such plots are designed to allow the separation of different excitation
mechanisms, thereby providing useful insights into the physics of a particular
object. In the present paper they are used to separate pure shocks from pure
precursors and shocks with ``full'' (photon-bounded) precursors, as well as
discriminating between results expected for shocks of different velocities and
different values of $B/n^{1/2}$. Hence, these diagrams provide an independent
means of investigating the consistency of shocks with the observations. The
ratios are taken from Table~\ref{tab_ratios} and represent the following
distinct regions:
(1)~the {\it Central} elongated emission-line region;
(2)~the total {\it Tail} region, representing the tidal ``tail'' structure
extending away from the central region;
(3)~the three individual ``tail'' sections {\it A}, {\it B} and {\it C}, for
which fluxes of some of the brighter lines were obtained.

\begin{figure*}
\begin{minipage}{\linewidth}
\centering\epsfig{file=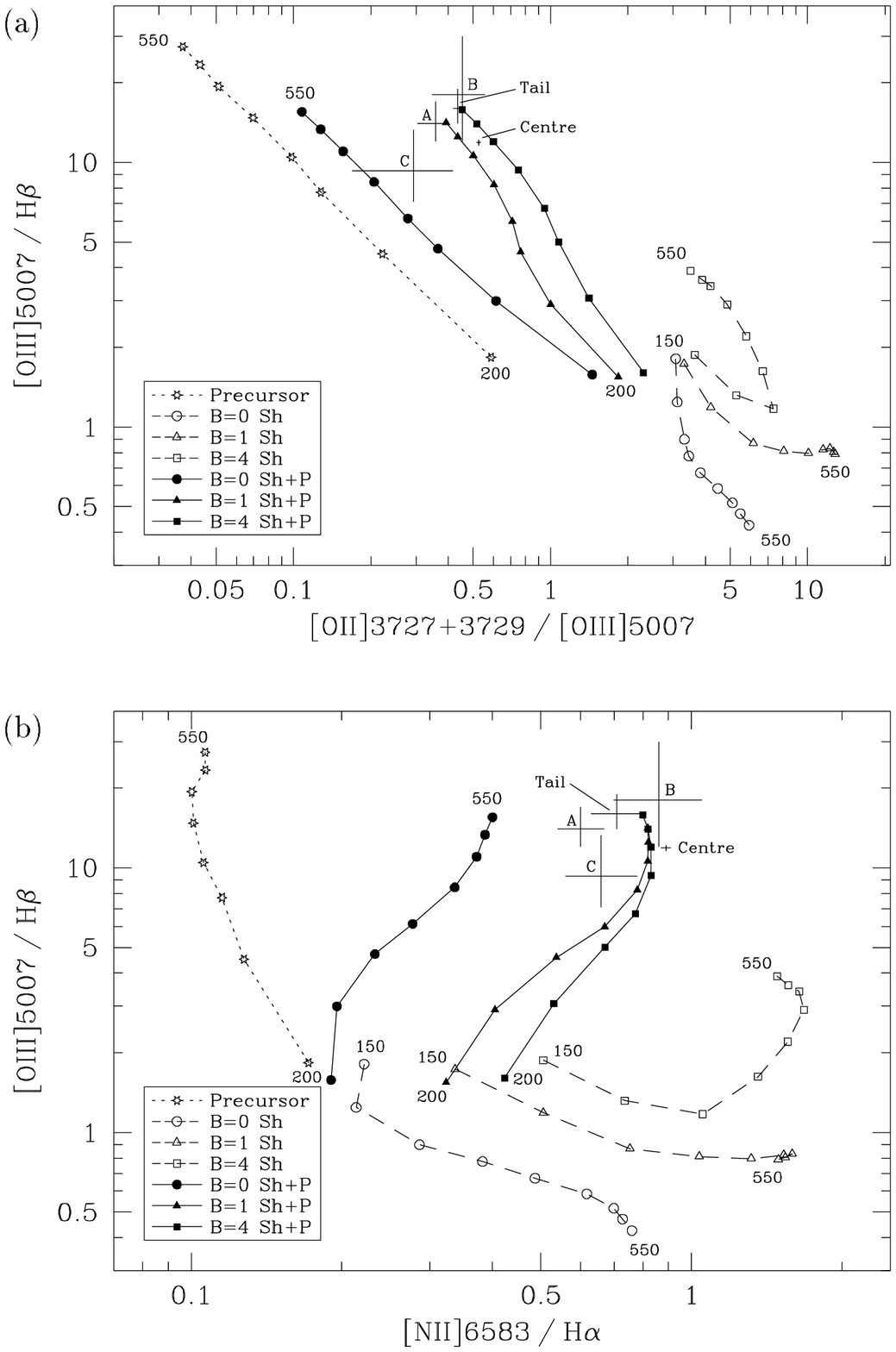,
	height=530pt}
\end{minipage}
\caption{
Diagnostic ratios of:
$(a)$ [OIII]$\lambda$5007/H$\beta$ vs. [OII]$\lambda\lambda$3727+3729/[OIII]$\lambda$5007;
$(b)$ [OIII]$\lambda$5007/H$\beta$ vs. [NII]$\lambda$6583/H$\alpha$. All the
model curves extend to 550$\kms$, with velocities plotted at increments of
50$\kms$. Loci for pure precursor emission commence at 200$\kms$, while those
involving shocks commence at 150$\kms$. Note that the value of the
$B$-parameter significantly affects emission from the shocked region, but not
from the precursor material.
}
\label{fig_diagnostic_1}
\end{figure*}

\begin{figure*}
\begin{minipage}{\linewidth}
\centering\epsfig{file=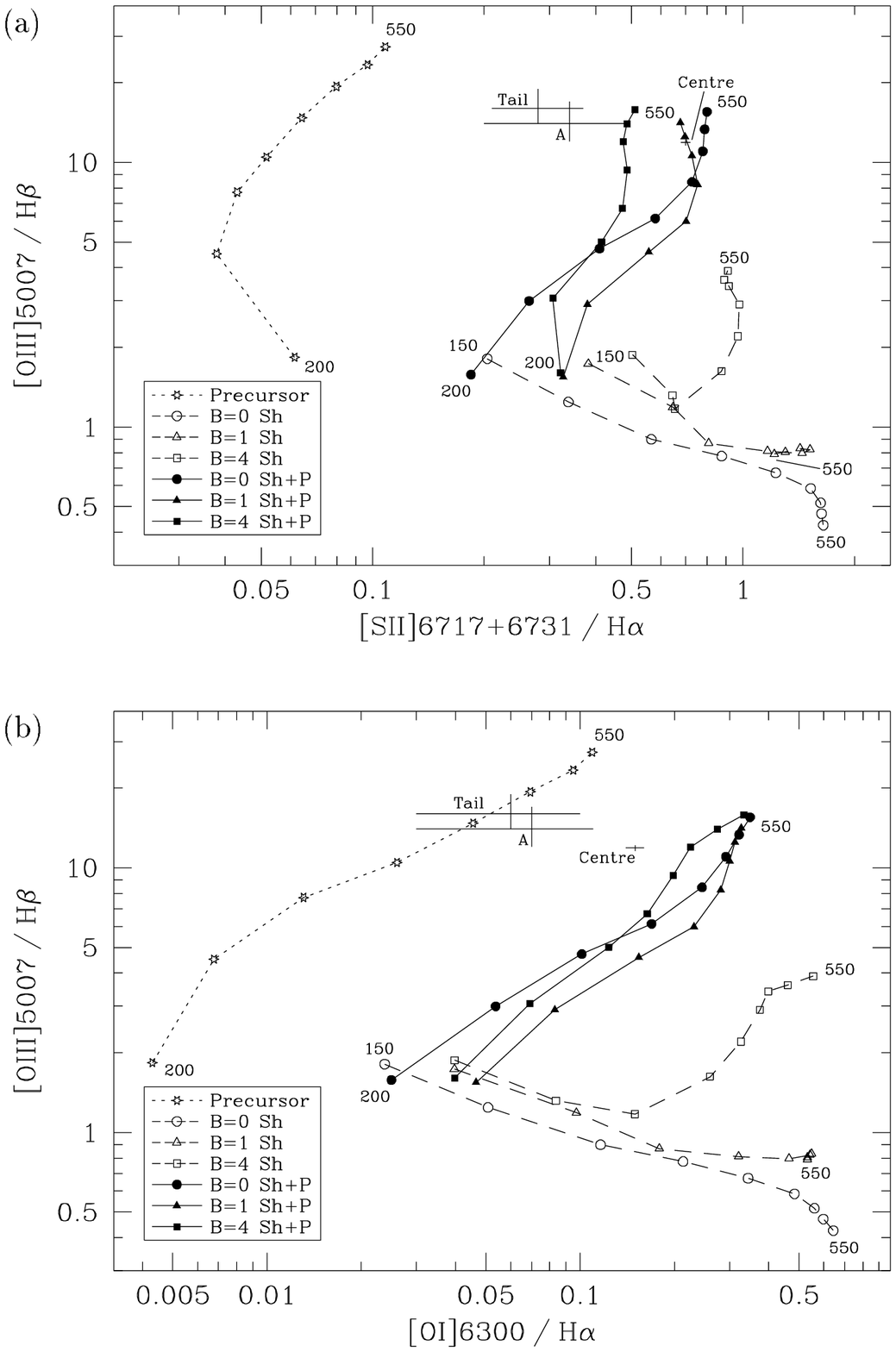,
	height=530pt}
\end{minipage}
\caption{
Diagnostic ratios of:
$(a)$ [OIII]$\lambda$5007/H$\beta$ vs. [SII]$\lambda\lambda$6717+6731/H$\alpha$;
$(b)$ [OIII]$\lambda$5007/H$\beta$ vs. [OI]$\lambda$6300/H$\alpha$.
The theoretical shock loci are plotted the same way as in
Figure~\ref{fig_diagnostic_1}.
}
\label{fig_diagnostic_2}
\end{figure*}

\begin{figure*}
\begin{minipage}{\linewidth}
\centering\epsfig{file=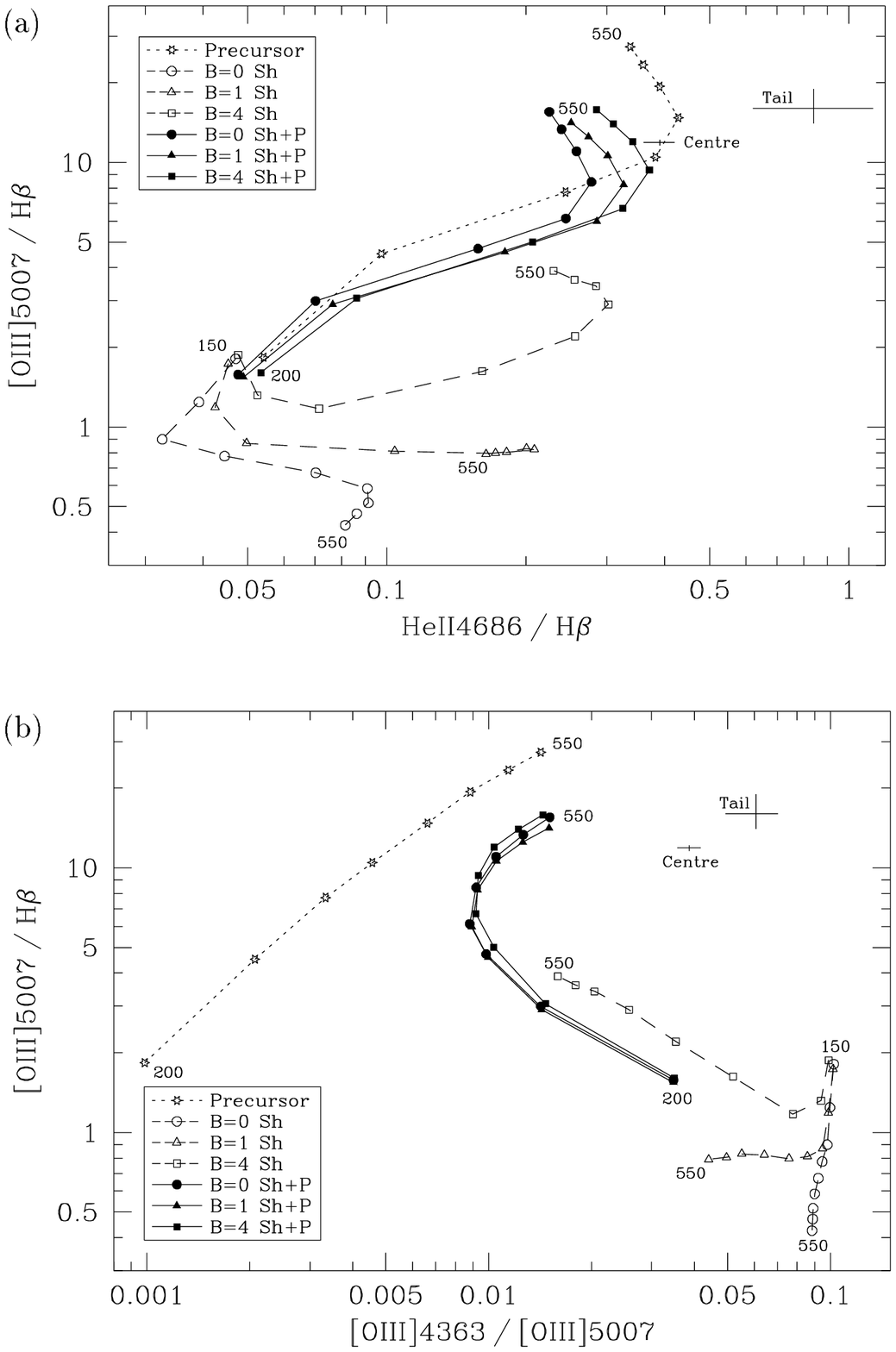,
	height=530pt}
\end{minipage}
\caption{
Diagnostic ratios of:
$(a)$ [OIII]$\lambda$5007/H$\beta$ vs. HeII$\lambda$4686/H$\beta$;
$(b)$ [OIII]$\lambda$5007/H$\beta$ vs. [OIII]$\lambda\lambda$4363/5007.
The theoretical shock loci are plotted the same way as in
Figure~\ref{fig_diagnostic_1}.
}
\label{fig_diagnostic_3}
\end{figure*}

The relevant ratios and \hbox{1-$\sigma$} error limits for these regions are
presented on the diagnostic line ratio plots in
Figures~\ref{fig_diagnostic_1}~--~\ref{fig_diagnostic_3}, together with the
loci expected from shock models. The \hbox{1-$\sigma$} limits are purely
indicative of the statistical errors of the measured fluxes; they do not
include effects such as line ratio variations across individual regions. Note
that fluxes for some of the fainter lines could not be obtained in all three
tail regions {\it A}, {\it B} and {\it C}, thus they are omitted from some of
the plots. A comparison of the observed data with the models allows a number of
physical properties to be inferred about the gas and allows a good test of the
self-consistency of the the shock model, which we discuss here for various line
ratios.

[OIII]$\lambda$5007/H$\beta$ vs. [OII]$\lambda\lambda$3727/[OIII]$\lambda$5007
(Fig. \ref{fig_diagnostic_1}$a$). ---
The relatively high values of the ratio of [OIII]$\lambda$5007/H$\beta$ imply
the presence of photoionized precursors, since the shocked material alone
produces comparatively weak [OIII]$\lambda$5007 emission. The observed data
points are consistent with combined shock and precursor emission, with shock
velocities in the range $\sim 400 - 550\kms$ and a magnetic parameter value
$B/n^{1/2} \sim 1\,\mu$G$\,$cm$^{3/2}$. However, the points for {\it A}
and {\it C} lie somewhat closer to ``pure'' precursor emission than the region
{\it B} and the Central region; it should also be noted that uncertainties in
the reddening correction could either increase or decrease the
[OII]$\lambda$3727/[OIII]$\lambda$5007 ratio by $\sim 30 - 50\%$.

[OIII]$\lambda$5007/H$\beta$ vs. [NII]$\lambda$6583/H$\alpha$
(Fig.~\ref{fig_diagnostic_1}$b$). ---
The positions of the [NII]$\lambda$6583/H$\alpha$ points with respect to the
model curves are remarkably consistent with those in
Figure~\ref{fig_diagnostic_1}$a$, even though the ratio of
[NII]$\lambda$6583/H$\alpha$ is observationally independent of the
[OII]$\lambda$3727/[OIII]$\lambda$5007 ratio. This agreement offers
strong support for the feasibility of the shock model. In particular,
Figure~\ref{fig_diagnostic_1}$b$ also suggests that the emission from the gas
is due to shocks combined with photon-bounded precursors, and that the magnetic
field parameter is likely to be $B/n^{1/2} \sim 1\,\mu$G$\,$cm$^{3/2}$ rather
than zero. This is a typical value of the magnetic parameter expected for ISM
gas
(Dopita \& Sutherland 1995).
\nocite{Dopita.1995.ApJ.455.468}
%
%
It is also interesting to note that the agreement with
Figure~\ref{fig_diagnostic_1}$a$ implies that nitrogen abundances are not
likely to be substantially different from Solar values (\ie within a factor
$2 - 3$).

[OIII]$\lambda$5007/H$\beta$ vs. [SII]$\lambda\lambda$6717+6731/H$\alpha$
(Fig. \ref{fig_diagnostic_2}$a$). ---
The tail regions {\it B} and {\it C} are absent from this figure, since the
[SII] emission from these regions is of too low S/N to be useful. Thus the only
fluxes plotted here are from the Central region, from the tail region {\it A},
and from the summed spectrum of all three tail regions. The results apparent on
this diagram are very similar to those of Figure~\ref{fig_diagnostic_1}, \ie
the ratios are dominated by emission from radiation-bounded precursors
photoionized by shocks of velocities in the range $450 - 550\kms$, with a
magnetic parameter $B/n^{1/2} \sim 1\,\mu$G$\,$cm$^{3/2}$. In particular, the
emission from the tail region {\it A} (and the tail as a whole) appears
dominated by emission from precursors rather than post-shock material, while
the central region is, as before, consistent with post-shock emission together
with kinematically associated photoionized precursors. We also note that the
relative consistency of these results with those of
Figure~\ref{fig_diagnostic_1} implies that additional effects (such as
collisional de-excitation) are unimportant, thus the gas is predominantly in
the low-density limit ($n_e \lesssim 10^3\,$cm$^{-3}$).

[OIII]$\lambda$5007/H$\beta$ vs. [OI]$\lambda$6300/H$\alpha$
(Fig.~\ref{fig_diagnostic_2}$b$). ---
These ratios are also consistent with ionization involving shocks of velocities
$\gtrsim 400\kms$. As in Figure~\ref{fig_diagnostic_2}$a$, the
[OI]$\lambda$6300 emission in the tail is relatively weak and is primarily from
region {\it A}, the location of which suggests emission dominated by ``pure''
precursor material. The Central region lies substantially closer to the loci
comprising combined emission from precursors and shocks with velocities
$\sim 400 - 450\kms$, and a weak value for the magnetic parameter. It should be
noted that despite the importance of [OI]$\lambda$6300 as a diagnostic tool for
discriminating between optically thick and thin photoionization conditions, its
modelling entails substantial uncertainties. For example, the formation of
molecules in the partially ionized region can lower the electron temperature
and significantly reduce the [OI] emission
(Dopita \& Sutherland 1996),
\nocite{Dopita.1996.ApJS.102.161}
%
%
and this is not currently accounted for in the MAPPINGS~II code. Predicted
model values of the [OI]$\lambda$6300 line emission tend to vary by a factor
$\sim 2 - 3$ between different ionization codes
(Ferland et~al. 1995),
\nocite{Ferland.1995.AEL.83}
%
%
a substantially larger variation than for the higher ionization states of
oxygen. The discrepancy between the MAPPINGS~II models and the observations is
also of this order, thus it is possible that incorporating this effect into the
models would make Figure~\ref{fig_diagnostic_2}$b$ more consistent with
Figure~\ref{fig_diagnostic_2}$a$, although such modifications are well beyond
the scope of the present work. Despite this uncertainty, however, the results
from the [OI]$\lambda$6300/H$\alpha$ ratio are still remarkably consistent with
those obtained from the other ratios.

[OIII]$\lambda$5007/H$\beta$ vs. HeII$\lambda$4686/H$\beta$
(Fig.~\ref{fig_diagnostic_3}$a$). ---
The HeII$\lambda$4686 flux for the central region agrees well with the scenario
inferred from the other ratios, \ie combined emission from shocks of velocity
$\sim 400 - 500\kms$ with radiation-bounded precursors, and magnetic field
parameters of order $\gtrsim 1\,\mu$G$\,$cm$^{3/2}$. The value of
HeII$\lambda$4686/H$\beta$ for the tail region is approximately consistent with
pure precursors photoionized by shocks of velocity $\sim 400 - 450\kms$; the
value of the ratio is somewhat higher (about a factor of 2) than expected from
the models, but it should also be noted that the observational uncertainty for
this ratio in the tail region is comparatively large. Thus, these results are
overall approximately consistent with those obtained from
Figures~\ref{fig_diagnostic_1} and \ref{fig_diagnostic_2}.

[OIII]$\lambda$5007/H$\beta$ vs. [OIII]$\lambda$4363/[OIII]$\lambda$5007
(Fig. \ref{fig_diagnostic_3}$b$). ---
Upon comparison with the data presented in
Dopita \& Sutherland (1995)
\nocite{Dopita.1995.ApJ.455.468}
%
%
it is clear that the electron temperature for the gas in this galaxy is among
the highest seen in radio galaxy emission line regions. It is likely that
uncertainties in the underlying continuum features have led to an
overestimation of the [OIII]$\lambda$4363 flux; however, the largest possible
error introduced in this way corresponds to an overestimation of the
[OIII]$\lambda$4363 flux by a factor of $\sim 2-3$. Reducing the observed
[OIII]$\lambda$4363 flux by this amount would still provide good agreement, not
only with the fluxes predicted by the shock model, but also with the particular
scenario inferred from the other diagrams --- combined precursor and post-shock
emission from shocks of velocities $\sim 400 - 550\kms$. By comparison, nuclear
photoionization models
typically under-predict this ratio by factors $\gtrsim 50 - 100$
(\eg Stasi{\'n}ska 1984; Tadhunter, Robinson \& Morganti 1989).
\nocite{Stasinska.1984.AA.135.341,
	Tadhunter.1989.EAG.293}
%
%
Such discrepancies can be reduced to some extent through
the use of lower than solar abundances
(Ferland \& Mushotzky 1982; Ferland \& Netzer 1983)
\nocite{Ferland.1982.ApJ.262.564,
	Ferland.1983.ApJ.264.105}
%
or the introduction of matter-bounded clouds having various distributions of
thickness
(Morganti et~al. 1991; Viegas \& Prieto 1992; Binette et~al. 1996).
\nocite{Morganti.1991.MNRAS.249.91,
	Viegas.1992.MNRAS.258.483,
	Binette.1996.AA.312.365}
%
%
However, in order to compare those results with the models presented here, we
would need to introduce similar free parameters, such as different
metallicities, and varying distributions of precursor optical depths or shock
velocities along the line-of-sight, and such modifications are outside the
scope of the present work.

In summary, the following scenario can be inferred from the comparison of three
observationally independent line ratios, [OIII]$\lambda$5007/H$\beta$,
[NII]$\lambda$6583/H$\alpha$ and [OII]$\lambda$3727/[OIII]$\lambda$5007: the
ratios in the central region and the tail region {\it B} are consistent with
combined emission from shocks with velocities in the range $400 - 550\kms$ and
magnetic field parameters of order $B/n^{1/2} \sim 1\,\mu$G$\,$cm$^{3/2}$,
together with radiation-bounded precursors photoionized by these shocks. The
ratios in the tail clouds {\it A} and {\it C} tend to be more indicative of
emission dominated by precursors, photoionized by shocks with velocities of
order $400 - 550\kms$. The fact that consistency between the observations and
the models can be achieved in these ratios, without invoking additional free
parameters such as ``partial'' precursors (\ie matter-bounded), density
differences, or abundance variations, is a very strong point in support of the
viability of auto-ionizing shocks as an ionization mechanism for the EELR.

The examination of two additional line ratios, [OI]$\lambda$6300/H$\alpha$ and
[SII]$\lambda\lambda$6717+6731/H$\alpha$, suggests that the temperature in the
extended weakly ionized region may be somewhat lower than expected, possibly as
a result of the formation of molecular species. This is the only modification
required that is not possible to incorporate into the current model. It is also
apparent that the gas is predominantly in the low-density limit. Finally, the
shock model accounts for the ratio [OIII]$\lambda$4363/[OIII]$\lambda$5007 to
within a factor of $2 - 3$ (or closer, if uncertainties due to the underlying
continuum are taken into consideration), and this is more than an order of
magnitude superior to the estimates generally obtained from nuclear
photoionization models.

\section{Discussion}

We now consider the feasibility of different possible physical scenarios
inferred from the observations, in terms of the spatial geometry, dynamics and
ionizing mechanisms operating in the EELR. For simplicity, we discuss
separately the two most plausible sources of ionizing photons: the AGN or
shocks. While it is possible that both of these mechanisms are operating to
some extent, perhaps even to different degrees in different parts of the EELR,
in the present work our aim is rather to investigate the {\it dominant}
mechanism responsible for the bulk of the EELR energetics across the largest
spatial scales. This greatly reduces the number of possible combinations of
free parameters, thereby allowing more useful constraints to be obtained on the
physics of the line-emitting gas.

\subsection{Structure and Geometry of the Tail EELR\label{tail_sec5.1}}

Firstly, the structure and physical properties of the three clouds in the tail
region are investigated using the ionization rates presented in
Table~\ref{tab_Nion}, and the observed projected spatial distribution of the
clouds on the sky. This enables constraints to be placed on their intrinsic
distances from the nucleus, together with related physical properties such as
the covering fraction and volume filling factor of line-emitting gas in these
clouds.

We consider a scenario in which the three tail clouds are primarily
photoionized by radiation escaping from the central region. This is motivated
by the relatively low linewidths in the tail clouds: for example, the velocity
width across all of cloud {\it C} is $\lesssim 150\kms$, much too small to be
directly associated with the shock velocities required to ionize the gas
(unless all the shocks are highly inclined to the line of sight, which is very
improbable). We later discuss the possible effect of supplementary ionization,
if shocks with velocities $\sim 200 - 300\kms$ are present within the tail
clouds {\it A} or {\it B}.

If the primary source of ionizing photons is considered to lie towards the
centre of the galaxy, this has the additional advantage that the derived tail
cloud properties are relatively insensitive to whether the radiation is
produced by the AGN or a collection of shocks throughout the central region. We
note, however, that the inferred properties of the central EELR depend strongly
upon the scenario invoked for production of the ionizing radiation, and in the
next sub-section this is used to investigate the relative viability of shocks
or the AGN as the primary source of ionizing photons.

We denote by $N_{\rm tot}$ the total rate of ionizing photons produced by the
radiation source; a fraction ${\mathcal F}_{\rm cen}$ of these photons are
absorbed by the entire central EELR, while the remaining flux of escaping
ionizing photons is $N_{\rm esc} = (1 - {\mathcal F}_{\rm cen})\, N_{\rm tot}$.
For each tail cloud we use the values of $N_{\rm ion}$ presented in
Table~\ref{tab_Nion} to constrain $N_{\rm esc}$. The covering fraction of
material in each tail cloud is represented by ${\mathcal F}$, where
${\mathcal F}=1$ indicates total absorption of all incident ionizing radiation
by the cloud. The observed projected separation $\Delta$ between each tail
cloud and the nucleus is related to their intrinsic separation $R$ by the angle
$i$, defined by $R \propto \Delta / \cos i$ (thus $i=0\arcdeg$ corresponds to a
geometry where the line between the nucleus and the tail cloud lies in the
plane of the sky).

\begin{deluxetable}{lccc}
\tablewidth{0pt}
\tablecaption{\label{tab_Nesc}
	Flux of Escaping Ionizing Radiation.}
\tablehead{
		& Distance $\Delta$ & $N'_{\rm esc}$		& $R'$	\\
		& (arcsec)	& (s$^{-1}$)			& (kpc)
}
\startdata
Tail {\it A}	& $\phn6\arcsec$ & $5.8 \times 10^{52}$		& 22.2	\nl
Tail {\it B}	& $10\arcsec$	& $1.1 \times 10^{53}$		& 26.5	\nl
Tail {\it C}	& $15\arcsec$	& $2.3 \times 10^{53}$		& 27.9	\nl
\enddata
\end{deluxetable}

In Table~\ref{tab_Nesc} we present
$N'_{\rm esc} = N_{\rm esc} {\mathcal F} \cos^2 i$, which for each tail cloud
corresponds to the inferred total rate of ionizing photons escaping from the
ionizing source, required to produce the observed values of $N_{\rm ion}$
(taking into account the parameters ${\mathcal F}$ and $i$ for each cloud). The
projected distance between each cloud and the nucleus, together with the
estimated spatial extent of each cloud, are used to determine the solid angle
subtended by each cloud at the nucleus. We also define a quantity $R'$,
representing the distance between the nucleus and tail clouds {\it A} and
{\it B} that would be required to produce the same value of $N'_{\rm esc}$ as
calculated for tail cloud {\it C\,}:
\begin{displaymath}
R' = R \times \left({ {\mathcal F}_C \cos^2 i_C \over
					{\mathcal F}}\right)^{1 \over 2}
\end{displaymath}
Thus, $R'$ parameterizes the intrinsic distance $R$ between a tail cloud and
the nucleus, together with the covering fraction ${\mathcal F}$ of the cloud,
relative to ${\mathcal F}_C$ (the covering fraction of cloud {\it C\,}) and
$i_C$, the angle between the vector from the nucleus to cloud {\it C} and the
plane of the sky. Cloud {\it C} is chosen as a reference because it is the
furthest from the nucleus, thereby minimizing the effects of uncertainties in
estimates of its spatial extent and its separation from the nucleus.
Although some uncertainties are involved in these derivations, for example
approximations concerning the precise three-dimensional structure of each tail
cloud, it is still possible to investigate the contribution of effects such as
the inclination and covering fraction on our estimates of $N_{\rm esc}$, under
the following three physically different scenarios.

Firstly, if the covering fractions ${\mathcal F}$ are approximately the same
for each tail cloud, then the values of $N'_{\rm esc}$ and $R'$ suggest a
geometry where tail cloud {\it C} is closest to the plane of the sky (with
respect to the nucleus) and tail cloud {\it A} is furthest out of the plane of
the sky. More specifically, if the vector between the nucleus and tail cloud
{\it C} lies approximately in the plane of the sky
\hbox{($i \lesssim 20\arcdeg$)}, then \hbox{$i \sim 50\arcdeg$} for the vector
between the nucleus and tail cloud {\it B}, and \hbox{$i \sim 65\arcdeg$} for
the vector to cloud {\it A}. This would imply that all three tail clouds are at
almost the same intrinsic distance from the nucleus ($R \sim 25\,$kpc), moving
on an orbit with low ellipticity. Interestingly, this could be consistent with
the observed kinematics: cloud {\it C} shows the greatest velocity offset with
respect to the galaxy nucleus, cloud {\it A} displays a velocity almost
identical to systemic (as would be expected if its tangential orbital velocity
is strongly projected), and cloud {\it B} displays an intermediate velocity.
One weakness of this scenario is that it does not readily account for the
observed velocity split in tail cloud {\it B}; we later discuss the kinematics
in more detail and consider possible extensions to this scheme.

A second possibility is that the covering fraction of the tail clouds increases
as a function of radius. If all three tail clouds have the same value of $i$,
then the covering fractions for clouds {\it B} and {\it C} relative to cloud
{\it A} are ${\mathcal F}_B \sim 1.8 {\mathcal F}_A$ and
${\mathcal F}_C \sim 3.8 {\mathcal F}_A$ respectively. This might possibly be
accounted for in terms of radial differences in the external medium around each
cloud (as discussed in
Binette et~al. 1996, or Prieto et~al. 1993).
\nocite{Binette.1996.AA.312.365,
	Prieto.1993.MNRAS.263.10}
%
%
For example, if the covering fraction is directly proportional to the volume
filling factor, and if this depends in turn upon the external pressure, then a
radial decrease of pressure as $\sim r^{-1.4}$ may produce the observed change.
However, this is inconsistent with the radial profiles of $r^{-2}$ or steeper
generally obtained for ellipticals and similar systems
(Kochanek 1994b; de~Zeeuw \& Franx 1991
\nocite{Kochanek.1994.ApJ.436.56,
	deZeeuw.1991.ARAA.29.239}
%
%
and other references therein). Furthermore, there exist no detailed physical
models describing such effects produced by the surrounding medium upon the
volume filling factor of material within a line-emitting region, hence this
scenario is difficult to quantify further.

Finally, in the context of photoionization by the AGN, the differences in
$N_{\rm esc}$ might be explained by different amounts of radiation reaching the
three clouds, as a result of anisotropic emission. For example, cloud {\it A}
may lie further toward the edge of an ionizing ``cone'', while clouds {\it B}
and {\it C} may be closer to its axis, thus more of the ionizing radiation from
the AGN is obscured towards cloud {\it A} compared with {\it B} or {\it C}.
Unfortunately, in this scenario the possible combinations of cloud positions
and cone opening angles and relative degrees of obscuration are almost entirely
arbitrary --- the observed structure may be matched with several completely
different cloud positioning configurations and cone angles, which do not
readily allow independent verification through other observational properties.
Perhaps more importantly, a highly anisotropic ionization cone is not easily
reconciled with the uniform radial flux distribution of the central EELR.

We adopt here the simplest parameterization for the observed differences in
$N_{\rm esc}$, assuming that they are primarily due to changes in a single
variable, either $i$ or ${\mathcal F}$. If we consider $i$ to be the free
variable, setting ${\mathcal F}$ to be the same for all three clouds, then the
tabulated values of $R'$ (or $N'_{\rm esc}$) imply that the clouds are all at
approximately the same distance from the nucleus ($R'/\cos i_C \sim 25\,$kpc).
Hence, the clouds would all be in approximately the same environment, thereby
providing an independent physical justification as to why ${\mathcal F}$ should
be similar for all the clouds. The alternative possibility, considering
${\mathcal F}$ to be the dominant free parameter and fixing $i$ to be the same
for each cloud, is inherently more complex since it demands additional
assumptions concerning differences in the environment around each cloud, as
well as a detailed description of how changes in the environment translate into
differences in covering fraction.

Thus we consider the different values in $N'_{\rm esc}$ to arise primarily from
different values of $i$, implying that the tail clouds are at similar distances
from the nucleus (small variations may be accounted for by differences in
${\mathcal F}$, related to the detailed shapes of the clouds). This scenario is
appealing for several reasons: (1) the number of free parameters is minimal;
(2) the inferred values of $i$ are comparatively insensitive to the spatial
structure of the source of ionizing photons (\ie AGN or shocks), provided that
they originate closer to the nucleus than the tail cloud itself; (3) the
derived tail cloud locations yield a dynamical model that can be rigorously
tested through a comparison with the kinematic structure.

\subsection{Energetics of the Ionizing Source}

The tail cloud fluxes provide constraints upon global properties of the central
EELR, such as the fraction of ionizing photons that are absorbed, given a
specific scenario for the ionization mechanism. In this section we examine the
energetics required to account for the emission from the central EELR and the
tail clouds, in the context of either shocks or AGN photoionization. We also
extrapolate the ionizing spectra into the optical regime and compare the
expected fluxes with the observed low-dispersion spectra.

Since the tail cloud {\it C} is considered to be entirely photoionized by
external radiation, its value of $N'_{\rm esc}$ can be used to infer directly
the flux of ionizing photons, $N_{\rm esc} (\nu > \nu_0)$, escaping from the
ionizing source:
\begin{displaymath}
	N_{\rm esc} (\nu > \nu_0)
		= {2.3 \times 10^{53} \over {\mathcal F}_C \cos^2 i_C}
					{\hspace{1em} \rm s^{-1}}
\end{displaymath}
Here $\nu_0 = 13.6\,$eV is the threshold ionization frequency for Hydrogen,
which we use to obtain the total ionizing flux from ionization models (thus
allowing optical fluxes to be computed and compared with the observed spectra).
We consider the general geometry of ionizing photons escaping isotropically
from the central region, before discussing other possibilities such as
selective absorption towards certain directions (or possible contributions from
local shocks in the tail clouds).

\subsubsection{Photoionization by the AGN}

We first discuss the EELR energetics in the context of ``unified schemes'', in
which all the ionizing photons are considered to originate from the AGN. The
total rate of ionizing photons absorbed by the central EELR is given by the
corresponding value of $N_{\rm ion}$ in Table~\ref{tab_Nion}; an upper limit on
the total rate of escaping ionizing photons, implied by the tail cloud fluxes,
is given by the highest value of $N'_{\rm esc}$ (that of tail cloud {\it C\,})
in Table~\ref{tab_Nesc}. Hence we obtain directly an upper limit on
${\mathcal F}_{\rm cen}$, the only unknown parameters being the values of
${\mathcal F}$ and $i$ for tail cloud {\it C}:
\begin{displaymath}
{\mathcal F}_{\rm cen} =
		\left(1 + {0.733 \over {{\mathcal F}_C \cos^2 i_C}}\right)^{-1}
\end{displaymath}
If the AGN produces a power-law spectrum of the form
\hbox{$F_\nu \propto \nu^\alpha$}
(\eg Robinson et~al. 1987)
\nocite{Robinson.1987.MNRAS.227.97}
%
%
then the value of $N_{\rm ion}$ for the central EELR, combined with the upper
limit on ${\mathcal F}_{\rm cen}$, yield direct lower limits on the expected
flux from the AGN in the optical regime. For example, setting $\alpha = -1.5$
implies a flux density
$F_\lambda \gtrsim 5.3 \times 10^{-16}\,$erg$\,$s$^{-1}\,$cm$^{-2}\,$\AA$^{-1}$
at $\lambda \sim 4000\,$\AA, which is at least an order of magnitude higher
than the observed upper limit on the possible featureless component in the
nuclear continuum (the absorption feature at $\lambda = 3935\,$\AA\ in the
spectrum of the central 3$\times$3\arcsec, presented in
Figure~\ref{fig_spec_1}$b$, displays a minimum flux level of
$F_\lambda = 5.1 \times 10^{-17}\,$erg$\,$s$^{-1}\,$cm$^{-2}$\AA$^{-1}$). The
most likely interpretation of this deficit involves a high degree of
obscuration toward the AGN, which would be in agreement with the lack of
observed nuclear broad emission lines in this object and also generally
consistent with the ``unified'' scenario of obscuration along our line-of-sight
towards the nuclei of narrow-line radio galaxies.

The difficulty with accounting for the EELR energetics predominantly by
radiation from the AGN is that the central EELR displays no morphological signs
of biconical photoionization, but instead has a very uniform radial
distribution, and is somewhat extended along a position angle apparently
unassociated with the radio axis (offset by $\sim 40\arcdeg$). This morphology
is also evident in the narrow-band image previously published by
Danziger \& Focardi (1988).
\nocite{Danziger.1988.CFCG.133}
%
%
Furthermore, the radial density distribution required to account for the tail
fluxes is too flat compared with physically reasonable models of elliptical
galaxies, as already discussed in sub-section~\ref{tail_sec5.1}. Thus, we are
motivated to investigate another possible mechanism for producing the bulk of
the ionizing radiation, by means of shocks within the EELR.

\subsubsection{Ionization by Shocks}

In this sub-section we discuss the feasibility of accounting for the EELR
energy budget primarily in terms of shocks, rather than radiation from the AGN.
Once again, tail cloud {\it C} is considered to be photoionized, thus we assume
that its recombination line luminosity follows the same scaling relationships
as derived for precursor emission by
Dopita \& Sutherland (1996).
\nocite{Dopita.1996.ApJS.102.161}
%

We consider first a geometry where all the ionizing radiation is produced by
shocks occurring in the central EELR (neglecting for the moment the possibility
of second-order contributions from lower-velocity shocks within the tail clouds
{\it A} or {\it B\,}). The physical significance of the quantity
${\mathcal F}_{\rm cen}$ in this scenario is that it represents the fraction of
ionizing photons absorbed by precursor material (of density $n_{\rm cen}$)
directly associated with the shocks in the central EELR. For simplicity, the
shocks are all assumed to correspond to a single velocity $V_S$; thus we may
directly apply the luminosity scaling relationships for shocks and precursors
given in Equations~(3.4) and (4.4) respectively of
Dopita \& Sutherland (1996).
\nocite{Dopita.1996.ApJS.102.161}
%
%
(together with ratios presented in their Tables~8 and 10 to scale the
luminosities from H$\beta$ to H$\alpha$, in order to afford a comparison with
the higher S/N observed H$\alpha$ fluxes). This yields the expected total
observable H$\alpha$ flux produced by the shocks and precursors in the central
EELR, necessary to produce the required number of escaping ionizing photons to
photoionize the tail cloud {\it C} (having a density $n_C$):
%
%
\begin{eqnarray*}
\lefteqn{F_{\rm H\alpha,{\rm cen}} = { 1.68 \times 10^{-14} \over
	{\mathcal F}_C \cos^2 i_C (1 - {\mathcal F}_{\rm cen}) }
		\times \left( n_{\rm cen} \over n_C\right)  }	\\
	& & \times \left[0.787 \left({V_S \over 100\kms}\right)^{0.13} +
		{\mathcal F}_{\rm cen}\right]
			{\rm erg\,s^{-1}cm^{-2}}
\end{eqnarray*}
Although the scenario considered here is comparatively simple, involving shocks
of a single velocity, and assuming precursors of uniform density, it accounts
remarkably well for the total observed EELR luminosities. For example, if
${\mathcal F}_C \cos^2 i_C \sim 1$ (thus ${\mathcal F}_{\rm cen} \sim 0.6$),
then the expected value of $F_{\rm H\alpha,{\rm cen}}$, as required to
photoionize cloud {\it C} through photons escaping from shocks, agrees well
with the observed central EELR H$\alpha$ flux of
$2.29 \times 10^{-14}\,$erg\,s$^{-1}$\,cm$^{-2}$ (see Table~\ref{tab_fluxes}).
The agreement can be improved if the average density of precursor material in
the central region is a factor of $\sim 2 - 3$ times lower than that of cloud
{\it C}, which would physically correspond to the scenario of material in the
central region being more disrupted and diffuse than the cooler gas in the tail
region. Furthermore, if ${\mathcal F}_C \cos^2 i_C \ll 1$ then larger density
differences can also be accommodated.

The expected continuum flux density in the optical regime from the shocks and
precursors in the central region can be obtained directly from the MAPPINGS~II
model output files, and compared with the observed limits on non-stellar
emission. We find that at $\lambda \sim 4000\,$\AA, the expected total
underlying continuum emission from shocks and precursors in the central EELR
(taking $V_S = 500\,$km\,s$^{-1}$ and ${\mathcal F}_{\rm cen} \sim 0.6$) is 
$F_\lambda \sim 1.5 \times 10^{-17}\,$erg$\,$s$^{-1}\,$cm$^{-2}$\AA$^{-1}$; see
also Figure~5 in
Sutherland et~al. (1993).
\nocite{Sutherland.1993.ApJ.414.510}
%
%
This is well within the observed upper limit of
$F_\lambda = 2 \times 10^{-16}\,$erg$\,$s$^{-1}\,$cm$^{-2}$\AA$^{-1}$ in the
absorption minimum at $\lambda = 3935\,$\AA\ in the total spectrum of the
central EELR (Fig.~\ref{fig_spec_1}$a$), thus the required shock luminosity
does not contradict the observational limits upon the non-stellar continuum.

Some shocks may also possibly be present within the tail region itself; in
particular, the double velocity system in tail cloud {\it B} is suggestive of
two kinematically and spatially related emission-line systems. The geometry of
tail cloud {\it C} with respect to {\it B} is such that it should intercept
$\sim 15\,$\% of any ionizing radiation escaping from cloud {\it B}; a similar
argument to the one above reveals that ionizing photons escaping from any
shocks in cloud {\it B} can account only for $\lesssim 10\,$\% of the observed
emission from cloud {\it C}. Thus, the ionizing photons from any shocks present
within the tail region are more likely to be absorbed by their associated
precursors instead of escaping to ionize other parts of the EELR. This would
suggest that the volume filling factor of material in the tail clouds is closer
to unity than in the central EELR, again consistent with a scenario in which
the central material is strongly disrupted, while the tail is still relatively
quiescent.

\subsection{Shocks: Physical Characteristics}

We have so far demonstrated that the EELR morphology, excitation, kinematics
and luminosities can be successfully accounted for if the energy budget is
dominated by autoionizing shocks, predominantly occurring within the central
EELR, and that this scenario requires less unknown parameters than the
alternative scheme of photoionization by radiation from the AGN. Here we
discuss the viability of the shock hypothesis in terms of the overall
physical properties required of the shocks and their precursors.

The spatial structure of the gas is clearly an important consideration in
discussing the occurrence of shocks. We point out that the volume filling
factor $f$ of the line-emitting gas, as commonly inferred from the
density-sensitive [SII] doublet, is only a reliable density indicator {\it if}
the emission is dominated by material with approximately uniform density. Thus,
the values of $n_e$ derived from the [SII]
ratio, combined with the constraints on $n_e f^{1/2}$ derived from the
ionization rates (Table~\ref{tab_Nion}) would suggest $f \sim 5 \times 10^{-6}$
for the emission-line gas in the central region. While this is generally
consistent with values inferred by other authors
(\eg van~Breugel et~al. 1985, 1986; Heckman et~al. 1989),
\nocite{vanBreugel.1985.ApJ.290.496,
	vanBreugel.1986.ApJ.311.58,
	Heckman.1989.ApJ.338.48}
%
%
this interpretation is valid only if the emission is entirely from material
photoionized by an AGN.

On the other hand, as we have already alluded to in
Sub-section~\ref{sec3.4_PhysCond}, if the emission is primarily related to
shocks then the [SII] ratios would be dominated by emission from the post-shock
cooling region, where the typical densities are two to three orders of
magnitude higher than in the precursor zone. For example, shocks with
$V_S \sim 450 - 500\,$km\,s$^{-1}$ and
$B/n^{1/2} \sim 1 - 2\,\mu$G$\,$cm$^{3/2}$ have post-shock regions
$\sim 200 -400$ times denser than their precursors, implying precursor
densities $n_{\rm cen} \lesssim 0.1 - 1\,$cm$^{-3}$ (if the density value for
the central region derived in Sub-section~\ref{sec3.4_PhysCond} applies
primarily to post-shock material). This would suggest values of $f$ within an
order of magnitude from unity for the unshocked precursor material in the
central EELR, thereby profoundly altering the physical description of the gas.

The radiative cooling timescale for material in the post-shock region is
typically $\sim 10^4\,$yr, too short for a single shock event to account for
the ionization of the gas during the lifetime of the radio source (generally
inferred to be $\sim 10^6 - 10^7\,$yr, for powerful radio galaxies,
\eg Alexander \& Leahy 1987).
\nocite{Alexander.1987.MNRAS.225.1}
%
%
One possibility for continuous deposition of energy into shocks is a direct
association between the radio jet and the line-emitting plasma, as postulated
for Centaurus~A
(Sutherland et~al. 1993),
\nocite{Sutherland.1993.ApJ.414.510}
%
%
and also more recently for PKS~2250$-$41
(Clark et~al. 1997).
\nocite{Clark.1997.MNRAS.inpress}
%
%
However, PKS~2356$-$61 displays very little association between the EELR and
the radio axis. In fact, the EELR morphology and kinematic structure are more
consistent with accretion onto the host galaxy; thus we propose for this object
a scenario involving shock formation that is driven by collisions between
infalling gas, and material that is already settling in the host galaxy.

Using the combined expressions for precursor and shock emission as a function
of shock velocity, together with the observed H$\alpha$ flux from the central
EELR, we find that the total surface area of shocks in the central EELR
required to produce the observed emission is:
%
%
\begin{eqnarray*}
\lefteqn{A_{\rm sh} = 1.44 \times 10^3 
	\left(n_{\rm cen} \over {\rm cm}^{-3}\right)^{-1}
	\left({V_S \over 100\kms}\right)^{-2.28}	}	\\
	& & \times \left[0.787 \left({V_S \over 100\kms}\right)^{0.13} +
		{\mathcal F}_{\rm cen}\right]^{-1} \hspace{1em} {\rm kpc}^2
\end{eqnarray*}
(where ${\mathcal F}_{\rm cen}$ represents, as before, the fraction of ionizing
radiation produced by shocks that is absorbed by precursors within the central
EELR). Given a precursor density $n_{\rm cen} \sim 0.1\,$cm$^{-3}$, with
${\mathcal F}_{\rm cen} \sim 0.6$, and shock velocities
$V_S = 500\,$km\,s$^{-1}$, the total mass flux through the shocks is
$\dot{M} \sim 1.26\,M_{\sun}\,$kpc$^{-2}\,$yr$^{-1}$. We note the possibility
of ``reprocessing'' of post-shock material by further shocks during the
lifetime of the source, in which case the mass flux through the shocks need not
necessarily be a direct indicator of the total mass of gas in the central EELR.

\subsection{Dynamics and Origin of the Gas}

The strong kinematic and spatial association between the clouds in the ``tail''
EELR suggests that they have a common dynamical history; here we discuss their
origin in the context of gas accretion by the host galaxy, as a result of an
interaction with a gas-rich companion. This scenario is favored on the basis
of several pieces of observational evidence: (1)~the presence of the H$\alpha$
object towards the north of the source; (2)~the faint bridge of emission-line
gas connecting the H$\alpha$ object with the host galaxy; (3)~the fact that
this faint bridge apparently lies on the same trajectory as the ``tail'' EELR
to the south of the host galaxy. Furthermore, we rule out the following
alternative explanations for the gas origin: (1)~ejection of gas along the
radio axis is inconsistent with the lack of any observed association between
the EELR and the radio axis; (2)~condensation from a cooling flow in the ICM is
not favored because of the comparatively high observed velocities, and also
the fact that the large-scale EELR is concentrated in a ``tail'', not a diffuse
halo around the host galaxy as expected for cooling flows
(Heckman et~al. 1989; Baum et~al. 1992; Fabian 1994).
\nocite{Heckman.1989.ApJ.338.48,
	Baum.1992.ApJ.389.208,
	Fabian.1994.ARAA.32.277}
%

Numerical studies of gas dynamics during galaxy encounters generally show a
tendency for the gas distribution to bifurcate into two components
(Combes et~al. 1990; Hernquist \& Mihos 1995; Barnes \& Hernquist 1991, 1996):
\nocite{Combes.1990.DIG.205,
	Hernquist.1995.ApJ.448.41,
	Barnes.1991.ApJ.370.L65,
	Barnes.1996.ApJ.471.115}
%
%
a compact, centrally condensed region, and extended ``tidal tails''. While the
detailed morphologies and dynamics clearly depend upon the interaction
geometry, this general bifurcation is physically understood as a direct
consequence of the decreased dynamical timescale towards the centre of the
accreting galaxy, the gas having been driven into the central region through
gravitational torquing in response to the tidal forces
(Binney \& Tremaine 1987; Barnes \& Hernquist 1992):
\nocite{Binney.1987.GalDyn,
	Barnes.1992.ARAA.30.705}
%
%
material within the inner few kpc experiences several orbital crossings and
begins to virialize while material extended on scales $\gtrsim 10\,$kpc is
still undergoing infall or large-scale orbital motions.

The ``settling'' of gas towards the centres of elliptical galaxies has also
been alluded to in several observational studies
(\eg Bertola \& Bettoni 1988; M{\"o}llenhoff \& Bender 1990),
\nocite{Bertola.1988.ApJ.329.102,
	Mollenhoff.1990.DIG.274}
%
%
together with more detailed simulations of the {\it kinematic} evolution of the
gas
(Habe \& Ikeuchi 1985, 1988; Varnas et~al. 1987).
\nocite{Habe.1985.ApJ.289.540,
	Habe.1988.ApJ.326.84,
	Varnas.1987.ApJ.313.69}
%
%
A typical feature of such simulations is the presence of large velocity shears
in the gas, allowing the formation of shocks within the settling gas itself;
shocks can also be expected as a result of the interaction between infalling
gas and the settling gas
(Combes et~al. 1988; Athanassoula 1992).
\nocite{Combes.1988.AA.203.L9,
	Athanassoula.1992.MNRAS.259.345}
%
%
Interestingly, similar results have been obtained in a different context,
namely the disruption and orbital motion of stellar debris near a black hole
(Kochanek 1994a; Lee et~al. 1996).
\nocite{Kochanek.1994.ApJ.422.508,
	Lee.1996.ApJ.464.131}
%
%
In such cases, the simulations clearly reveal collisions between material
settling in the potential well and streams that are still in the process of
accretion, with the resulting shocks contributing significantly to a loss
of angular momentum of the infalling streams.

Preliminary three-dimensional hydrodynamic simulations of interactions between
a small gas-rich companion galaxy and a large elliptical
(Koekemoer 1996),
\nocite{Koekemoer.1996.PhD}
%
%
using parameters appropriate to those inferred for PKS~2356$-$61, indicate that
infalling material is likely to interact directly with the settling gas for a
comparatively wide range of encounter geometries and galaxy/halo
configurations, and that such interactions persist over timescales
$\sim 10^7 - 10^8\,$yr. Direct connections between shock formation and tidal
accretion have also been proposed elsewhere in the literature, for example the
peculiar spiral galaxy NGC~4438
(Combes et~al. 1998; Keel \& Wehrle 1993; Kenney et~al. 1995).
\nocite{Combes.1988.AA.203.L9,
	Keel.1993.AJ.106.236,
	Kenney.1995.ApJ.438.135}
%
%
Substantially more sophisticated modelling will be needed before more
quantitative statements can be made concerning the properties of any shocks
formed during the accretion process, in particular detailed three-dimensional
descriptions of the cooling and fragmentation processes occurring behind the
shocks, as well as realistic models of the sizes and spatial distributions of
gas clouds within the tidal streams. However, the results obtained to date
suggest that shocks can play a significant role in galaxy interactions, with
their energetic importance depending primarily upon the velocities involved.

Thus, the formation of shocks within the central EELR can be understood as
being driven both by the dynamics of the settling gas itself, as well as by
the additional infalling material. Since the timescale associated with galaxy
interactions is of the same order as the lifetime of the radio source, the
EELR energetics can feasibly be powered in this scenario for much of the
lifetime of the source. The energy supply for the EELR is ultimately related to
the gravitational potential energy released by the infalling material, the
kinetic energy associated with the initial relative velocity between the
galaxies, and the total mass of gas accreted by the host galaxy.

\section{Conclusions}

We have presented results from detailed optical longslit spectroscopy of the
EELR in the nearby powerful radio galaxy PKS~2356$-$61, providing complete
three-dimensional spectroscopic datacubes that sample the entire spatial
distribution of line-emitting gas. We have carried out an extensive
investigation of the observed physical properties of the gas, comparing these
with results expected from the MAPPINGS~II shock ionization code, and we have
found that the excitation, kinematics and energetics of the gas are entirely
consistent with a scenario involving auto-ionizing shocks as the dominant
ionization mechanism.

We find that the EELR is separated into two kinematically and spatially
distinct components: a central region, having a typical velocity width
$\sim 300 - 400\kms$, and a tail region extending away from the centre,
having much lower line-of-sight velocity widths ($\sim 120 - 150\kms$) and
displaying more internal velocity structure than the central region. We have
also discovered a faint H$\alpha$-emitting galaxy located $\sim 60\,$kpc toward
the north of the host galaxy of PKS~2356$-$61, exhibiting two elongated streams
of emission apparently indicating a dynamical association between the two
galaxies.

The highly detailed nature of our data has allowed us to test in several
complementary ways the feasibility of the proposition that the gas energetics
may be powered by auto-ionizing shocks. Specifically, we have compared the
EELR line-of-sight velocity widths with both the emission-line fluxes and
excitation relationships expected from the MAPPINGS~II models, and we have also
presented MAPPINGS~II loci on diagrams of observed diagnostic line ratios
plotted against one another. These tests repeatedly yield similar results: the
properties of the central region are consistent with emission from shocks with
intrinsic velocities in the range $\sim 400 - 550\kms$, together with their
associated precursors, while emission from the tail clouds are more indicative
of ``pure'' precursor material, photoionized by shocks of these velocities but
not kinematically associated with them. This is further confirmed by an
examination of the EELR energy budget, in which we find that a total covering
fraction ${\mathcal F}_{\rm cen} \sim 0.6$ for the precursor material in the
central region is sufficient to account for the properties of the central EELR,
while at the same time providing agreement with the observed fluxes of the
photoionized clouds in the tail region.

Although the alternative hypothesis of photoionization by radiation from an
active nucleus cannot be completely ruled out in this object, we find that the
radially uniform morphology of the central EELR cannot be readily reconciled
with the anisotropic ionization ``cone'' anticipated in such a model.
Furthermore, nuclear photoionization requires additional arbitrarily free
parameters, in particular the spectral shape and total flux of the ionizing
continuum, together with specific amounts of obscuration required to explain
the fluxes in the tail clouds. On the other hand, the shock model can account
for all the observed EELR properties by means of a single physical process, and
allows direct observational verification of all the primary observable
quantities --- shock velocities, line fluxes and excitation --- thereby greatly
reducing the number of arbitrary parameters.

The morphology of the EELR appears completely unrelated to the radio axis, and
we find it unlikely that interactions between the radio plasma and the gas play
a significant role in this source. Rather, we propose that the formation of the
shocks is essentially related to gravitationally-driven kinetic energy
deposited by gas that is being accreted into the host galaxy from a companion
object. While detailed three-dimensional radiative, non-equilibrium,
magneto-hydrodynamic modelling would be required in order to determine the
precise properties and lifetimes of shocks formed during such a process, the
observed velocities correspond well to those required for gas accretion to take
place during such interactions. Furthermore, the approximate interaction
timescale in the case of PKS~2356$-$61 is in the range $\sim 10^7 - 10^8\,$yr,
allowing for the possibility of a continual supply of shocks from fresh
infalling material during the lifetime of the radio source.

Interactions between galaxies have long been considered to play a role in
triggering nuclear activity as a result of gas accretion, particularly in the
context of powerful radio galaxies. Although the processes involved in such
interactions are likely to be quite complex, the possibility that shocks can
form as a result, at least in some cases, and dominate the EELR energetics, is
highly appealing in terms of accounting for all the observed properties of the
extended line-emitting gas by a single physical process. Clearly, similarly
detailed observations of other radio galaxies will be required in order to
investigate the importance of this process in the general context of the
formation and evolution of active galaxies.

\acknowledgements
{\begin{center}
\bf Acknowledgements
\end{center}}
We are grateful to the Anglo-Australian Observatory and ATAC for the support of
this project in the form of generous allocations of observing time, and we
thank the staff at the telescope for support during the observations. We thank
Mike Dopita and Ralph Sutherland for making available to us the MAPPINGS~II
code, which we used to calculate the model results presented in this paper.
We would also like to thank the referee, Dr.~Patrick McCarthy, for useful
comments which helped to improve the paper. A.M.K. acknowledges financial
support during the course of this work in the form of an Australian National
University Postgraduate Research Scholarship.

%
%

\cleardoublepage

\setcounter{figure}{1}

\begin{figure*}
\caption{
(colour plate).
Several views of the [OIII]$\lambda$5007 velocity datacube for PKS~2356$-$61,
from different angles. The blue and green axes correspond to Right Ascension
and Declination, with pixel dimensions of 1\arcsec\ and 0\farcs77 respectively.
The pixel aspect ratio is such that the spatial image scale is the same along
both axes. The red axis represents velocity, with increasing systemic velocity
towards the right in all frames. The velocity scale is $\sim 45\kms$/pixel and
the total velocity range is $\sim 1800\kms$. The colour scale corresponds to
pixel flux (the brightest emission is coloured red, decreasing to yellow, green
and blue). The flux scale is logarithmic, normalised to the rms noise level of
the faintest detected emission
($\sigma_0 \sim 1 \times
	10^{-18}\,$erg$\,$s$^{-1}\,$cm$^{-2}\,$arcsec$^{-2}\,$\AA$^{-1}$);
values of 1$\sigma_0$, 10$\sigma_0$, 100$\sigma_0$ and $\gtrsim 1000\sigma_0$
are indicated by dark blue, blue-green, yellow-orange and red, respectively.
}
\label{fig_cube_avs}
\end{figure*}

\setcounter{figure}{3}

\begin{figure*}
\caption{
(colour plate).
Spatial variation of emission-line flux ratios across PKS~2356$-$61:
$(a)$~[OIII]$\lambda$5007/H$\beta$; $(b)$~[NII]$\lambda$6583/H$\alpha$;
$(c)$~[SII]$\lambda\lambda$6716+6731/H$\alpha$; 
$(d)$~[SII]$\lambda\lambda$6716/6731; $(e)$~H$\alpha$/H$\beta$.
Pixel scales and image orientations are as for Fig.~\ref{fig_img_cont}. Colour
indicates the ratio value, and ratios with relatively low S/N are shown with a
correspondingly lower intensity.
}
\label{fig_img_line}
\end{figure*}

\end{document}